\documentclass[11pt]{article}
\usepackage[T1]{fontenc}
\usepackage[utf8]{inputenc}
\usepackage{textcomp}
\usepackage{lmodern}
\usepackage{amsmath,amssymb}
\usepackage[margin=1in]{geometry}
\usepackage{microtype}
\usepackage{parskip}
\usepackage{xcolor}
\usepackage{graphicx}
\usepackage{longtable,booktabs,array}
\usepackage{calc}
\usepackage{caption}
\usepackage{framed}
\usepackage{authblk}
\usepackage{etoolbox}
\makeatletter
\patchcmd\longtable{\par}{\if@noskipsec\mbox{}\fi\par}{}{}
\makeatother
\usepackage{hyperref}
\usepackage{xurl}
\hypersetup{hidelinks,
  pdftitle={Networked wealth and mortality in the Agincourt Health and Demographic Surveillance System 2009-2018},
  pdfauthor={Keletso Makofane, Eric Tchetgen Tchetgen, Mary T. Bassett, Lisa F. Berkman}}

\title{Networked wealth and mortality in the Agincourt Health and Demographic Surveillance System, 2009--2018}

\author[1,*]{Keletso Makofane}
\author[1,2]{Eric Tchetgen Tchetgen}
\author[3]{Mary T. Bassett}
\author[4]{Lisa F. Berkman}
\affil[1]{Department of Biostatistics, Epidemiology, and Informatics, Perelman School of Medicine, University of Pennsylvania, Philadelphia, Pennsylvania}
\affil[2]{Department of Statistics, The Wharton School, University of Pennsylvania, Philadelphia, Pennsylvania}
\affil[3]{Harvard FXB Center for Health and Human Rights, TH Chan School of Public Health, Harvard University, Boston, Massachusetts}
\affil[4]{Harvard Center for Population and Development Studies, Harvard University, Cambridge, Massachusetts}
\affil[*]{Corresponding author: \href{mailto:keletso.makofane@gmail.com}{keletso.makofane@gmail.com}}

\date{}

\begin{document}
\maketitle

\begin{center}
\begin{minipage}{0.9\linewidth}
\small\itshape
This is a pre-copyedited, author-produced version of an article accepted for publication in the \emph{American Journal of Epidemiology} following peer review (manuscript AJE-00063-2023).
\end{minipage}
\end{center}

\begin{abstract}
\textbf{Background.} Though resources embedded in social networks appear to shape health, past studies have not assessed this relationship directly. Using data from Agincourt Health and Demographic Surveillance System in South Africa, we examined the spillover effect of household wealth on mortality in the family network.

\textbf{Methods.} We analysed cohort data from 2009 to 2018 to estimate the Average Total Effect (ATE) of household wealth on deaths, decomposing this into an Average Direct Effect (ADE) and Average Spillover Effect (ASE). We estimated these quantities separately for all deaths, deaths among children (\textless5 years old), premature deaths (among ages 15-59), and deaths among elders (60+).

\textbf{Results.} Over the study period, 8750 individuals died. Overall, a one standard deviation increase in household wealth led to an ATE of -0.61\% (95\% CI: -0.74 to -0.48, P\textless0.001), an ADE of -0.53\% (95\% CI: -0.63 to -0.42, P\textless0.001) and an ASE of -0.08\% (95\% CI: -0.18 to 0.01, P=0.075). For premature mortality, these effect estimates were -0.37\% (95\% CI: -0.47 to -0.26, P\textless0.001), -0.29\% (95\% CI: -0.37 to -0.21, P\textless0.001), and -0.08\% (95\% CI: -0.15 to 0.00, P=0.038), respectively.

\textbf{Conclusions.} Results provide evidence for the protective effect of household wealth on premature mortality across family relationships.


\end{abstract}

\hypertarget{introduction}{%
\section{Introduction}\label{introduction}}

In the context of widespread poverty and unemployment, South African families employ several strategies to distribute life-sustaining material resources among their kin. Households are ``stretched'' across space, allowing migrant laborers to remit income; they are structured around resources by, for instance, housing unemployed adults with recipients of government transfers; and they give resources to, and receive them from, family members' households, depending on ability and need.\textsuperscript{1--4} Though these strategies are likely to induce interdependence in the wealth and health of family members' households, epidemiologic studies commonly make the assumption that measures taken in distinct households are independent, or that they are dependent only through spatial proximity, missing valuable opportunities to understand and quantify the influence of network-embedded resources on population health.\textsuperscript{5}

There are several empirical literatures that indirectly investigate this causal relationship. We focus on the social capital and health and the social networks and health literatures. In the former, social capital is typically defined as resources ``accessed by individuals as a result of their membership of a network or group''\textsuperscript{6} and analyzed using scales that assess the conditions under which relationships permit the transmission of resources between individuals.\textsuperscript{6}~In the latter, studies usually ask participants to list important social contacts and answer questions about each listed contact. In analysis, researchers characterize the type and frequency of support received from contacts as well as the structure of the ``egocentric'' network that connects them to the participant and to each other.\textsuperscript{7}

Prior studies suggest that such wealth shapes health in South Africa. Social capital has been shown to correlate with better oral health, improved HIV outcomes, lower risk of tuberculosis infection, higher self-rated health, better subjective well-being, and better mental health.\textsuperscript{8--18} A recent social network study found positive associations between cognitive health and social support, including financial and instrumental support.\textsuperscript{19}

While they provide useful insights, the measures typically used in epidemiologic studies of social capital and social networks are tangentially related to the central ideas of ``network'' and ``resource''.\textsuperscript{20} The social capital literature prioritizes the measurement of attitudes, perceptions, cognitions, and behaviors over the measurement of the distribution and transmission of material resources.\textsuperscript{6} Owing to its roots in psychosocial research, the social networks literature tends to investigate intangible forms of support (e.g.~emotional support) over material forms (e.g.~financial support).\textsuperscript{7} While the former literature makes no attempt to measure network structure, the latter does, but from the narrow perspective of only one member of the (egocentric) network under study. In this case, when material forms of support (such as financial support) are investigated, they tend to be measured in terms of the frequency and not the volume or monetary value of the support. Crucially, support from social contacts is measured without assessing their own access to resources.\textsuperscript{7,20}

These are crucial omissions. In societies with extremely unequal distributions of wealth and income, such as South Africa, social network and social capital measures that do not account for access to resources likely conceal heterogeneity in the quantity and quality of support individuals receive from others, compromising inferences about the impact of that support on health. For instance, a measure of financial support that does not account for the distribution of wealth cannot distinguish between a person who receives weekly financial support from a wealthy family member and a person who reports weekly support from a poor one. A measure of social capital that does not account for networks, cannot distinguish between a transgender person who perceives a high level of trust among members of their predominantly cisgender community and a cisgender person who perceives a high level of trust in the same setting. Whereas the cisgender person might benefit from the trust, because of transphobia, a transgender person might find themselves excluded from trusting relationships and the resources they avail.

Directly measuring and analyzing sociocentric social networks has the potential to overcome this limitation and advance a causal understanding of interdependence. A major barrier to the application of these methods, though, is that sociocentric network data are notoriously challenging to collect. It is logistically difficult to make direct measurements on all named social contacts of a given study participant, and ethically challenging to collect data from participants that pertains to social contacts who have not consented to participate in the study. Furthermore, even when they are collected, social network data cannot be analyzed using standard causal inference methods, which rely on the assumption that measures on observational units (or groups of observational units) are independent.\textsuperscript{21,22}

Our study addresses these gaps. Applying auto-regression analysis, we assessed the relationship between death in one household and wealth, which we take as a measure of available material resources, among the households of family members. We establish a causal interpretation of regression results drawing on auto-g-computation -- a recently developed approach to evaluating network causal effects.\textsuperscript{23} We define the average total effect (ATE) of household wealth on household mortality, decomposing this effect into an average direct effect (ADE) -- a causal component due to the household's own increase in wealth, and an average spillover effect (ASE) -- the portion of the ATE due to the increase in the wealth available to the household through family connections.

We estimate these parameters using observational data from the Agincourt Health and Demographic Surveillance System (AHDSS). In this system, each individual's record is associated with unique identifiers for her mother, father, and spouse. These identifiers were used to construct a sociocentric family network where, in addition to ties indicating these relationships, we included ties that connect parents-in-law, siblings, grand-parents, and other relatives who are two-steps away in the family tree. We coarsened this network into one where nodes are households. We present estimates of household-level ADE, ASE, and ATE along with the conditions under which the estimates can be interpreted as causal effects.

\hypertarget{methods-1}{%
\section{Methods}\label{methods-1}}

\hypertarget{study-setting}{%
\subsection{Study setting}\label{study-setting}}

Located in a \(420\) \(\text{km}^{2}\) region of Bushbuckridge Municipality, Mpumalanga Province, South Africa, AHDSS is a surveillance platform that conducts an annual census of residents, collecting information on births, deaths, migrations, family relationships, and wealth. Mpumalanga province has high unemployment and poverty - a recent survey shows that whereas in South Africa, 18\% of households were estimated to be food insecure in 2019, in Mpumalanga, this figure was 24\%.\textsuperscript{24,25}. About one fifth of adults in Bushcbuckridge Municipality have no schooling, compared to 6\% nationally.\textsuperscript{26,27}

The AHDSS has covered all households in a defined area in this region since 1992. The area expanded in 2013 to include four additional villages. AHDSS data are organized into a database which associates individuals' records with identifiers for the households they were a part of in each year of the census, and identifiers for their parents and conjugal partners. Since AHDSS covers all households in a defined geographic area, and since the refusal rate is low, we take AHDSS data to be representative of the underlying population.

\hypertarget{household-network}{%
\subsection{Household network}\label{household-network}}

For each year in the study period (2009 -- 2018), we used AHDSS data to construct a network where nodes are households and ties are family relationships between households. In this network, a pair of households is connected if at least one individual in the first household is linked to at least one individual in the second. For two individuals' records to be linked, they should have been members of the same household for at least one observation in the period 1992 to 2018.

The network was constructed in three steps (see Box 1). The first step (Step A) was to construct a network among all individuals who resided in the AHDSS between 1992 and 2019. In this network, nodes are individuals and ties represent two kinds of family relationships: parent-child relationships and conjugal (formalized romantic) relationships. In the second step (Step B), ties were added between each pair of nodes that were connected to a common node. This would mean that, for instance, each person has a direct connection to her mother and daughter (because of Step A) as well as her grandmother and sister (because of Step B). In the third step (Step C), for each calendar year between 2009 and 2018, we coarsened this network into one where nodes are households (instead of individuals) and ties are family relationships between households. A pair of households is connected by a tie if at least one individual in the first household is related to at least one individual in the second in the network produced by Steps A and B. Since individuals can change households over time, Step C was conducted separately for each calendar year between 2009 and 2018, producing a longitudinal household network over that period.

\hypertarget{household-and-network-deaths}{%
\subsection{Household and network deaths}\label{household-and-network-deaths}}

The unit of analysis is the household. We say a household has experienced death in a particular year if at least one member died in that year. Based on the age of the decedent, we defined four measures of household death: overall, child (under age 5), adult (ages 15-59), and elder (ages 60+). Age ranges were chosen to be consistent with prior work on mortality in Agincourt.\textsuperscript{28} Network deaths for a given household, in a given year, were computed as the count of connected households that experienced death. Network child deaths, network adult deaths, and network elder deaths were defined as the counts of connected households which experienced the death of a child, adult, or elder, respectively.

\hypertarget{household-and-network-wealth}{%
\subsection{Household and network wealth}\label{household-and-network-wealth}}

Household wealth was measured by an asset index calculated using DHS methodology.\textsuperscript{29,30} In this approach, field workers make observations on each household with respect to the construction materials of the main dwelling (e.g.~brick vs.~mud); the type of toilet facilities (e.g.~in dwelling vs.~neighbors compound); sources of water (e.g.~inside tap vs.~well), sources of energy (e.g.~electricity vs.~paraffin), and ownership of modern assets (e.g.~fridge, microwave) as well as livestock.\textsuperscript{31} A household wealth score is then computed as a weighted sum of indicator variables related to these components. Measurements of household wealth were made every two years from 2001 to 2013, and annually thereafter. In this analysis, for the years 2010 and 2012, we used the value of household wealth measured in 2009 and 2011, respectively. Network wealth for a given household in a given year was calculated as the sum of household wealth among connected households during that year. By doing this, we made the simplifying assumption that all family relationships contribute equally to resources transmitted across households.

\begin{figure}[tbp]
\begin{framed}
\textbf{Box 1: Constructing AHDSS household network}

\centering
\includegraphics[width=0.55\linewidth]{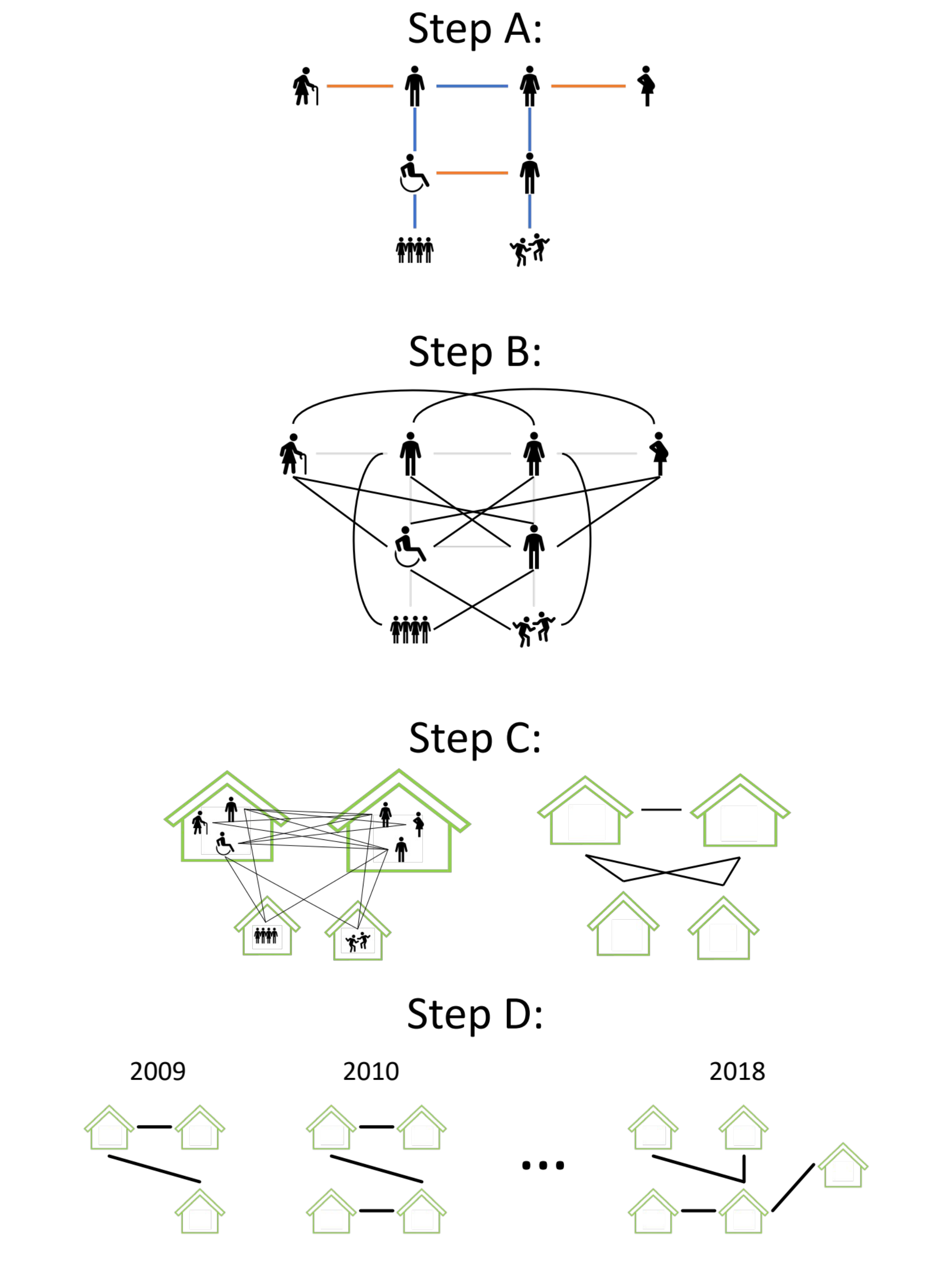}

\raggedright\small
\textbf{Step A}: A network was constructed among all individuals who resided in the AHDSS between 1992 and 2019. In this network, nodes are individuals and ties represent two kinds of family relationships: parent-child relationships (shown in orange) and conjugal relationships (shown in blue).

\textbf{Step B}: Ties were added (shown in black) between each pair of nodes that were already connected to a common node (shown in grey).

\textbf{Step C}: Using this network of individuals, we constructed a network where nodes are households and ties are family relationships between households. A pair of households is connected by a tie if at least one individual in the first household is related to at least one individual in the second. On the left, individuals are shown as icons contained by their households, shown in green. On the right, individuals are not shown. The households represented on the left are shown on the right with ties between households instead of individuals.

\textbf{Step D}: Finally, we repeated Step C for each year between 2009 and 2018. The structure of the household network changed over time since individuals can enter and leave the AHDSS and move between households within the AHDSS.
\end{framed}
\end{figure}

\hypertarget{causal-quantities-of-interest}{%
\subsection{Causal quantities of interest}\label{causal-quantities-of-interest}}

Following Tchetgen Tchetgen et al.,\textsuperscript{23} we calculated three causal estimands: the average direct effect (ADE), average spillover effect (ASE), and average total effect (ATE) (See appendix S1). We define a \(n \times 1\) vector of dichotomous potential outcomes \(\mathbf{Y}\left( \mathbf{a} \right)\) under \(n \times 1\) treatment vector \(\mathbf{a}\). The \(i^{th}\) element of this vector, \(Y_{i}\left( \mathbf{a} \right)\), is the potential outcome for household \(i\) under the level of household wealth dictated by \(\mathbf{a}\). The treatment for household \(i\) is \(a_{i}\) - the \(i^{th}\) element of \(\mathbf{a}\). \(\mathbf{L}\), \(\mathbf{Y}\), and \(\mathbf{A}\) are \(n\)-dimensional vectors of covariates, observed outcomes, and observed treatment, respectively. We make the assumptions of exchangeability and consistency, allowing us to identify the expected value of the mean of the counterfactual outcome as follows:

\[E\left\lbrack \mathbf{Y}\left( \mathbf{a} \right)|\mathbf{L} \right\rbrack = E\left\lbrack \mathbf{Y}\left( \mathbf{a} \right)|\mathbf{L},\mathbf{A} = \mathbf{a} \right\rbrack = E\left\lbrack \mathbf{Y}|\mathbf{L},\mathbf{A} = \mathbf{a} \right\rbrack\text{.}\]

For a focal household, we define the spillover effect as the change in the counterfactual risk of household death that results from intervening to increase every non-focal household's wealth by one standard deviation while keeping the focal household's wealth constant. ASE is the quantity we calculate by computing the spillover effect separately for each household and then taking an average over all households.

We define the direct effect as the change in the counterfactual risk of household death that results from intervening to increase a focal household's wealth by one standard deviation and to keep the wealth of all other households fixed. ADE is the average of this quantity across households.

ATE is defined as the change in the average counterfactual risk of household death that would result from an intervention to simultaneously increase wealth by one standard deviation in every household in the AHDSS. ATE is the sum of ADE and ASE.

ATE, ADE, and ASE are identified under the untestable assumptions that there is no unaccounted confounding, that the counterfactual outcomes that would obtain under the observed treatment regime are equal to the observed outcomes, and that the observed data arise from a semi-parametric model defined by the following two restrictions.

The first restriction is that the vector of household deaths is a realization of a conditional Markov random field defined by the following local Markov condition: household death in a focal household is independent of household death in any household it is not directly connected to in the network, conditional on the covariates and wealth of the focal household and the covariates, wealth and deaths of households connected to the focal household.\textsuperscript{23,32}

The second restriction is that we assume the following linear conditional mean auto-regression model holds:

\begin{equation}
\begin{aligned}
E\left\lbrack Y_{i,t + 1}|\mathbf{L}_{t},\mathbf{A}_{t},\mathbf{Y}_{t + 1} \right\rbrack = {} & \beta_{0} + \beta_{1}a_{i,t} + \beta_{2}l_{i,t} + \beta_{5}a_{i,t}l_{i,t} \\
 & + \sum_{j \in \mathcal{N}_{i,t},j \neq i}\left( \beta_{3}a_{j,t} + \beta_{4}l_{j,t} + \beta_{6}l_{i,t}a_{j,t} + \beta_{7}l_{i,t}l_{j,t} \right) \\
 & + \sum_{k \in \mathcal{N}_{i,t + 1},k \neq i}{\phi Y_{k,t + 1}}
\end{aligned}
\label{eq:1}
\end{equation}

where for household \(i\) in year \(t\), \(Y_{i,t}\) is household death, \(a_{i,t}\) is household wealth, \(l_{i,t}\) is a household covariate value, and \(\mathcal{N}_{i,t}\) is a set containing \(i\) and the indices of households connected to household \(i\) by family relationships. If there are \(n\) households in the network, \(\mathbf{L}_{t}\), \(\mathbf{A}_{t}\) and \(\mathbf{Y}_{t}\) are \(n\)-dimensional vectors containing covariates, wealth, and household death, respectively.

In the special case where \(\phi = 0\), the causal quantities of interest are given by:

\[
\begin{aligned}
ASE &= \beta_{3}\bar{d} + \beta_{6}\frac{1}{n}\sum_{i}{l_{i}d_{i}} \\
ADE &= \beta_{1} + \beta_{5}\bar{l}
\end{aligned}
\]

where \(d_{i}\) is the number of households connected to household \(i\) and \(\bar{d}\) is the average of \(d_{i}\) across households.

Though we show only one covariate for ease of presentation, in analyses, \(\mathbf{L}_{t}\) is a \(n \times k\) matrix containing: village, \# connected households, \# people in household, \# women in household, \# children in household, \# number of elders in household, \# people in connected households, \# women in connected households, \# children in connected households, \# number of elders in connected households.

\hypertarget{analysis}{%
\subsection{Analysis}\label{analysis}}

We conducted descriptive analyses of household and network deaths and covariates. To estimate ADE, ASE, and ATE, we first estimated the coefficients of the conditional mean model shown in Equation 1 using ordinary least squares, then plugged-in these estimates into the formulae for ASE and ADE shown above.

Obtaining unbiased coefficient estimates for the conditional mean model was complicated by possible network outcome dependence between deaths among connected households. The parameter \(\phi\) represents the extent of this dependence.

To obtain unbiased estimates of model coefficients when \(\phi \neq 0\) (i.e.~when there is network outcome dependence), the model should be fitted using a sub-sample of households such that no two households in the sub-sample are connected in the network.\textsuperscript{23} In the sub-sample, outcomes are conditionally independent (according to the local Markov condition). When \(\phi = 0\), the conditional mean model can be fitted using the entire sample.

To test for network dependence, we first fitted the conditional mean model using the entire sample and conducted a Wald test of the null hypothesis \(\phi = 0\). If we failed to find empirical evidence of network outcome dependence, we would proceeded with the analysis as if \(\phi = 0\).

To account for correlation across repeated measures within-households, we implemented the cluster bootstrap with household defining the independent unit. We re-sampled uniformly from the list of households 5000 times, each time estimating coefficients of Equation 1, and using these to compute ADE, ASE, and ATE.

\hypertarget{missing-data}{%
\subsection{Missing data}\label{missing-data}}

The household wealth index measure was missing for 9-34\% of households depending on the year. This induced missingness in the network wealth measure. All other variables were fully ascertained. To account for missing data, we implemented multiple imputation using a custom-built R-package based on the Multiple Imputation by Chained Equations (MICE) package.\textsuperscript{33} Household wealth was imputed using predictive mean matching. Network wealth was computed using imputed household wealth values. We created 8 complete datasets, each time iterating the custom-built MICE algorithm 60 times. Robust standard errors were calculated for each imputation. Coefficient estimates and standard errors were pooled across imputations using Rubin's rules.\textsuperscript{34} The imputation model produces valid inferences under the assumption that data were ``missing at random'' and that the functional forms of regressions used for imputation are correct.\textsuperscript{35} Further details are shown in Appendix S2.

\hypertarget{results-1}{%
\section{Results}\label{results-1}}

\hypertarget{descriptive-statistics}{%
\subsection{Descriptive statistics}\label{descriptive-statistics}}

From 2009 to 2018, the number of households in the Agincourt Health and Demographic Surveillance System grew from 16 725 to 21 291 (see Table 1 and Appendix S3) while the proportion of households that were not linked to others in the AHDSS rose from 30\% in 2009 to 36\% in 2018. These changes were largely driven by the inclusion of new villages to the AHDSS over time. Household size and gender and age composition did not appear to change substantially over this period. At the other extreme of connectivity, under 10\% of households were connected to six or more households in any given year.

From 2009 to 2018, 8750 people died. Consistent with the country-wide decline in mortality over this period,\textsuperscript{28,36,37} the annual overall proportion of households that experienced a death decreased from 5.4\% (904/16 275) in 2009 to 2.5\% (533/21 291) in 2018. Household deaths were dominated by working age adults since they comprise the largest age group. In 2018, elder individuals had the highest probability of death at 2.98\% (177/5 946) followed by working age adults (0.48\%; 322/66 580) and children (0.10\%; 44/41 203).

\emph{Table 1: Household Descriptive Statistics}

\begin{longtable}[]{@{}lrrr@{}}
\toprule\noalign{}
 & 2009 & 2014 & 2018 \\
\midrule\noalign{}
\endhead
\bottomrule\noalign{}
\endlastfoot
\multicolumn{4}{@{}l}{\textbf{Number of Households}} \\
N & 16275 (100.0\%) & 21527 (100.0\%) & 21291 (100.0\%) \\
\multicolumn{4}{@{}l}{\textbf{Village Type}} \\
Original Villages & 16275 (100.0\%) & 18144 (84.3\%) & 17924 (84.2\%) \\
Additional Villages & 0 (0.0\%) & 3383 (15.7\%) & 3367 (15.8\%) \\
\multicolumn{4}{@{}l}{\textbf{\# Connected Households}} \\
0 & 4769 (29.3\%) & 7840 (36.4\%) & 7645 (35.9\%) \\
1 & 3372 (20.7\%) & 4119 (19.1\%) & 4364 (20.5\%) \\
2 & 2734 (16.8\%) & 3107 (14.4\%) & 3190 (15.0\%) \\
3 & 1943 (11.9\%) & 2245 (10.4\%) & 2254 (10.6\%) \\
4 & 1332 (8.2\%) & 1560 (7.2\%) & 1491 (7.0\%) \\
5 & 799 (4.9\%) & 1045 (4.9\%) & 947 (4.4\%) \\
6+ & 1326 (8.1\%) & 1611 (7.5\%) & 1400 (6.6\%) \\
\multicolumn{4}{@{}l}{\textbf{\# Deaths in Household}} \\
0 & 15394 (94.6\%) & 20705 (96.2\%) & 20754 (97.5\%) \\
1 & 851 (5.2\%) & 795 (3.7\%) & 531 (2.5\%) \\
2 & 30 (0.2\%) & 26 (0.1\%) & 6 (0.0\%) \\
3+ & 0 (0.0\%) & 1 (0.0\%) & 0 (0.0\%) \\
\multicolumn{4}{@{}l}{\textbf{\# Household Members}} \\
1 & 1647 (10.1\%) & 2384 (11.1\%) & 2513 (11.8\%) \\
2-5 & 7380 (45.3\%) & 10075 (46.8\%) & 10053 (47.2\%) \\
6-10 & 5751 (35.3\%) & 7200 (33.4\%) & 7038 (33.1\%) \\
11-20 & 1458 (9.0\%) & 1834 (8.5\%) & 1657 (7.8\%) \\
21+ & 39 (0.2\%) & 34 (0.2\%) & 30 (0.1\%) \\
\multicolumn{4}{@{}l}{\textbf{\# Children (\textless5)}} \\
0 & 8234 (50.6\%) & 11731 (54.5\%) & 13190 (62.0\%) \\
1 & 5415 (33.3\%) & 6725 (31.2\%) & 6071 (28.5\%) \\
2-5 & 2607 (16.0\%) & 3058 (14.2\%) & 2028 (9.5\%) \\
6-10 & 19 (0.1\%) & 13 (0.1\%) & 2 (0.0\%) \\
\multicolumn{4}{@{}l}{\textbf{\# Adults (15-59)}} \\
0 & 606 (3.7\%) & 831 (3.9\%) & 854 (4.0\%) \\
1 & 2490 (15.3\%) & 3401 (15.8\%) & 3268 (15.3\%) \\
2-5 & 10695 (65.7\%) & 14099 (65.5\%) & 14095 (66.2\%) \\
6-10 & 2381 (14.6\%) & 3053 (14.2\%) & 2962 (13.9\%) \\
11-20 & 101 (0.6\%) & 142 (0.7\%) & 110 (0.5\%) \\
21+ & 2 (0.0\%) & 1 (0.0\%) & 2 (0.0\%) \\
\multicolumn{4}{@{}l}{\textbf{\# Elders (60+)}} \\
0 & 11510 (70.7\%) & 15038 (69.9\%) & 14258 (67.0\%) \\
1 & 3797 (23.3\%) & 5196 (24.1\%) & 5665 (26.6\%) \\
2 & 914 (5.6\%) & 1237 (5.7\%) & 1320 (6.2\%) \\
3+ & 54 (0.3\%) & 56 (0.3\%) & 48 (0.2\%) \\
\multicolumn{4}{@{}l}{\textbf{Household Wealth Index}} \\
Mean (sd) & 2.5 (0.46) & 2.7 (0.42) & 2.7 (0.39) \\
Median {[}Min - Max{]} & 2.6 {[}0.90 - 4.0{]} & 2.7 {[}0.90 - 4.0{]} & 2.7 {[}0.92 - 4.2{]} \\
Missing (\%) & 1498 (9.2\%) & 4340 (20.2\%) & 4761 (22.4\%) \\
\multicolumn{4}{@{}l}{\textbf{Network Wealth Index}} \\
Mean (sd) & 5.3 (5.7) & 5.1 (6.0) & 5.0 (5.9) \\
Median {[}Min - Max{]} & 3.8 {[}0.00 - 46{]} & 2.9 {[}0.00 - 52{]} & 2.9 {[}0.00 - 57{]} \\
Missing (\%) & 262 (1.6\%) & 744 (3.5\%) & 1025 (4.8\%) \\
\end{longtable}

\hypertarget{main-analysis}{%
\subsection{Main analysis}\label{main-analysis}}

We found support for our central hypothesis that wealth of related households is protective against mortality in a given household (see Figure 1 and Table S2 in Appendix S4). Overall, a 1-standard deviation increase in household wealth is associated with an ATE of -0.61\% (95\% CI: -0.74 to -0.48, P\textless0.001), an ADE of -0.53\% (95\% CI: -0.63 to -0.42, P\textless0.001), and an ASE of -0.08\% (95\% CI: -0.18 to 0.01, P=0.075).

This overall effect was largely driven by the deaths of working age adults. In this group, a 1-standard deviation increase in household wealth was associated with an ATE of -0.37\% (95\% CI: -0.47 to -0.26, P\textless0.001), an ADE of -0.29\% (95\% CI: -0.37 to -0.21, P\textless0.001), and an ASE of -0.08\% (95\% CI: -0.15 to 0.00, P=0.038).

\begin{figure}[tbp]
\centering
\includegraphics[width=\linewidth]{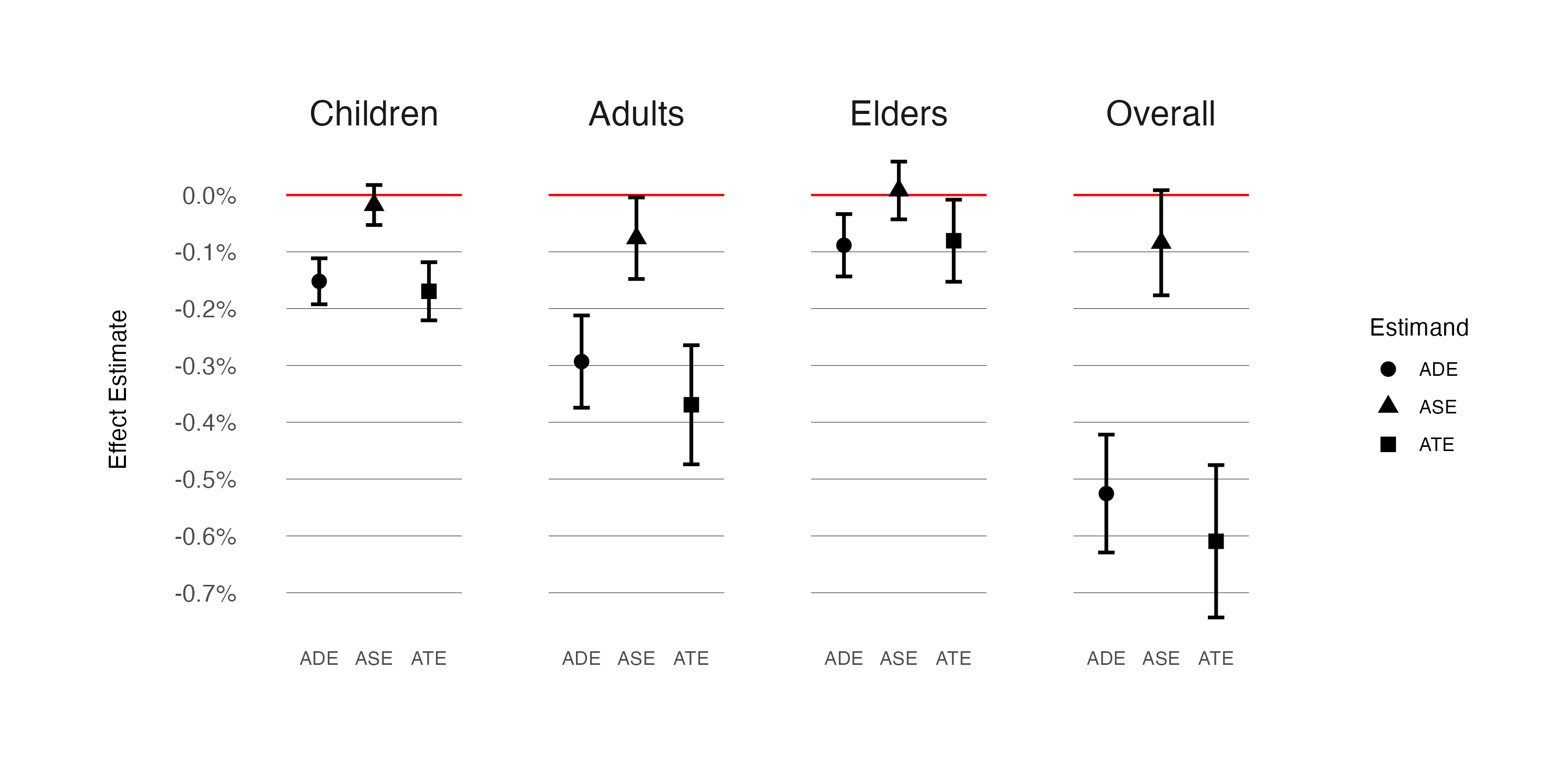}
\caption*{\textbf{Figure 1:} Direct and Spillover Effect of Family Network Wealth on Household Death. Agincourt Health and Demographic Surveillance System: 2009 to 2018. ADE: average direct effect; ASE: average spillover effect; ATE: average total effect.}
\end{figure}

Prior to conducting the main analysis, we failed to reject the null hypothesis of \(\phi = 0\) at the \(0.05\) significance level (\(P > 0.9\)) in the conditional mean model in Equation 1. i.e.~There was no statistical evidence that household death is correlated across connected households after adjusting for household wealth, network wealth, and covariates.

\hypertarget{discussion}{%
\section{Discussion}\label{discussion}}

Our study provides evidence for the protective effect of a household's wealth on deaths in that same household, both overall and within subgroups defined by age. There is moderate evidence for the protective effect against death in a given household, of wealth in connected households. This effect was strongest among working age adults, suggesting that material resources shape premature, and therefore preventable, mortality.

While a range of pathways potentially link household wealth to morbidity and mortality in family members' households, we suspect that our findings are explained by transfers of health-relevant material resources between connected households. In past studies, households have been shown to be embedded in local kinship-based networks of support. Members of extended family often live close to one another: ``people and food flow constantly between them, and labor, including childcare and supervision, is commonly shared or exchanged''.\textsuperscript{1} These patterns apply more broadly in South Africa -- in a nationally representative survey on giving in 2005, over half of the respondents reported having recently given money, goods, food, or other items to members of their family who lived in a different household.\textsuperscript{2} More recently, two in five South African households reported giving or receiving cash across family connections. This cash comprised as much as 40\% of household income.\textsuperscript{3,4} Private transfers have been shown to be progressive, moving from wealthier to poorer households, and from employed to unemployed people.\textsuperscript{4}

But along with family members, the South African government makes a substantial contribution to household income. Annually, the government distributes about 3\% of Gross Domestic Product through cash transfers, reaching about half of all households in South Africa.\textsuperscript{38,39} In 2019, transfers comprised 54\% of household income across Mpumalanga Province, with a further 22\% of income gained from remittances.\textsuperscript{40}

Agincourt has historically had a high unemployment rate, ranging from 63\% as measured in 2001 to 52\% as measured in 2011.\textsuperscript{41} The combination of private and state transfers, in the context of high unemployment, might explain why working-age adults were more sensitive to the level of wealth held in family networks than children and elders in our study. Whereas the government provides means-tested transfers to adults aged 65+ and to the primary caregivers of children, there is no transfer dedicated to unemployed adults unless they have a disability. It might be that in spending government transfers, households prioritize their intended beneficiaries -- children and elders, leaving unemployed working-age adults to be more dependent on other sources of household income, including private transfers. To advance an understanding of the mechanisms through which family members' resources act on health, it will be necessary to measure the form, frequency, and size of inter-household transfers and to understand how these transfers are applied to household expenses.

It is possible that the association between wealth in one household and death in another has more to do with the movement of sick people between connected households than it does with the movement of resources. Prior studies show that household structure and composition are dynamically determined by changing patterns of income, health, and wealth in South Africa. Members of extended family not only exchange resources between households, they form households around resources.\textsuperscript{1,42} Even if this were the case, our causal estimates would remain valid. The family connections that link households, however, would act as a conduit for redistributing resource needs rather one for redistributing resources.

It is not straightforward to generalize these findings to the national South African population.\textsuperscript{43} Though the age structure of AHDSS has been shown to be similar to that of the national South African census,\textsuperscript{44} it is possible that in the decade since this comparison was made, the demographic composition of AHDSS has deviated from that of South Africa. On the other hand, as Byass et al.~argue,\textsuperscript{45} mortality estimates in small-area studies tend to approximate national mortality trends.

Some limitations complicate the causal interpretation of our results. We assume that mortality is independent of wealth from prior years after adjusting for wealth in the year immediately preceding that in which we counted deaths. This assumption might not hold if household wealth in early life has a cumulative effect throughout the life course, leading to an impact on mortality in later life even after accounting for wealth in the intervening time. Studies have shown that socioeconomic status in early life can shape adult health in this way.\textsuperscript{46--48} Unfortunately, for any resident included in this analysis, the period over which household wealth was observed did not exceed 17 years.

It is possible that our effect estimates are biased by unmeasured confounding. Most crucially, if illness in the household decreases future household wealth through the loss of wages and through costs associated with caring for the ill, this could induce a negative relationship between wealth and mortality within and across households. But because of high unemployment and widespread poverty in Agincourt, and because of the welfare system in South Africa, it is not clear that illness reduces household wealth in subsequent time periods. Becoming ill enough to be incapable of working is an eligibility criterion for the state disability grant -- an unconditional cash transfer which, as Kelly et al.~recently argued, is a highly valued resource in households.\textsuperscript{49}

Finally, the measure of network wealth we used likely incorporates measurement error: We weighted family connections across households equally in terms of their potential to transmit resources across households. It is likely that closer relations (e.g.~a non-co-resident spouse) enable the transmission of larger quantities of resources than, more distant relationships (e.g. a father-in-law). In the extreme case, relatives most likely to give resources to a particular household might live in households that are not directly connected to them in our constructed network. Future studies should combine information on family structure with information on transfers between households.~

Through this study we make several substantive and methodological contributions. Prior investigations of material family support and health predominantly investigate the effect of support on health \emph{within} households.\textsuperscript{50,51} By contrast, in this study it is support between households whose health effects are assessed. This type of support has been studied when it occurs in the form of income remittances from migrants, but less so with respect to transfers between non-migrant kin.

Whereas most epidemiologic studies of social support or social networks rely on psychosocial and behavioral measures, we show that material support shapes mortality, and therefore health. To better understand the mechanisms through which social support influences health, it is crucial to incorporate explicit measures of material support in studies.

Past social network analyses have been limited by their inability to rule out homophily as an explanation for associations between ego health states and alter characteristics.\textsuperscript{52} We mitigated this bias by focusing on family relationships as these are largely established prior to health states and wealth. Finally, we demonstrated the feasibility of constructing population-based sociocentric social networks using routinely collected data and the feasibility of conducting epidemiologic analyses underpinned by an explicit framework for causal inference in the face of network spillover.

In summary, we found that material resources shape mortality across family networks, with the strongest effects seen on premature, and therefore potentially preventable, mortality. Social connections and the resources they array, therefore, are determinants of population health. Future studies should explore the multiple causal pathways through which these determinants act. To make this possible, it is crucial to cultivate sources of network data, including data that are collected passively, and to further develop methods for causal inference that accommodate causal spillover.

\hypertarget{declarations}{%
\section{Declarations}\label{declarations}}

\hypertarget{ethics-approval}{%
\subsection{Ethics approval}\label{ethics-approval}}

Ethical clearance for surveillance activities in the Agincourt HDSS was obtained from the Human Research Ethics Committee (Medical) of the University of the Witwatersrand (protocols M960720, M110138 and M180585).

\hypertarget{author-contributions}{%
\subsection{Author contributions}\label{author-contributions}}

KM designed and conducted the study, conducted literature review, and drafted the manuscript. ETT contributed to statistical methods. MB, LB, and ETT helped to interpret the findings and revised the manuscript critically for important intellectual content.

\hypertarget{data-availability}{%
\subsection{Data availability}\label{data-availability}}

The data underlying this article were provided by MRC/Wits Rural Public Health and Health Transitions Research Unit by permission. Data will be shared on request to the corresponding author with permission of MRC/Wits Rural Public Health and Health Transitions Research Unit.

\hypertarget{supplementary-data}{%
\subsection{Supplementary data}\label{supplementary-data}}

Supplementary materials (Appendices S1--S4) follow the references.

\hypertarget{funding}{%
\subsection{Funding}\label{funding}}

This work was supported by the National Institute on Ageing at the National Institutes of Health (grant number 1P01AG041710-01A1).

\hypertarget{conflict-of-interests}{%
\subsection{Conflict of interests}\label{conflict-of-interests}}

No potential conflict of interest was reported by the authors.

\hypertarget{acknowledgements}{%
\subsection{Acknowledgements}\label{acknowledgements}}

Re leboga ba Makofane -- dikgogwa tsa mohlakamotala, le ba Chere, Hlaele, Mathlape, Mocumi, le Ntebele; batsamai mmogo le rona.

\hypertarget{references}{%
\section{References}\label{references}}

1. Madhavan S, Mee P, Collinson M. Kinship in Practice: Spatial Distribution of Children's Kin Networks. \emph{Journal of Southern African Studies} {[}Internet{]}. 2014 Mar 4;\textbf{40}(2):401--418. Available from: \url{http://www.tandfonline.com/doi/abs/10.1080/03057070.2014.906211}

2. Everatt D, Habib A, Maharaj B, Nyar A. Patterns of Giving in South Africa. \emph{VOLUNTAS: International Journal of Voluntary and Nonprofit Organizations} {[}Internet{]}. 2005 Sep;\textbf{16}(3):275--291. Available from: \url{http://link.springer.com/10.1007/s11266-005-7725-z}

3. Posel D. Inter-household transfers in south africa: Prevalence, patterns and poverty. Cape Town; 2016.

4. Ssebagala RA. Plugging the Welfare Gap: The Role of Kinship Transfers in South Africa. \emph{The Journal of Development Studies} {[}Internet{]}. 2021 Jul 3;\textbf{57}(7):1154--1171. Available from: \url{https://www.tandfonline.com/doi/full/10.1080/00220388.2020.1850698}

5. Halloran ME, Hudgens MG. Dependent happenings: A recent methodological review. \emph{Current epidemiology reports}. 2016;\textbf{3}(4):297--305.

6. Kawachi I, Berkman LF. Social capital, social cohesion, and health. \emph{Social epidemiology}. 2014;290--319.

7. Berkman L, Krishna A. Social network epidemiology. \emph{Social Epidemiology}. 2014;234--289.

8. Olutola BG, Ayo-Yusuf OA. Socio-environmental factors associated with self-rated oral health in south africa: A multilevel effects model. \emph{International Journal of Environmental Research and Public Health} {[}Internet{]}. 2012 Oct;\textbf{9}(10):3465--3483. Available from: \url{https://www.ncbi.nlm.nih.gov/pmc/articles/PMC3509466/}

9. Mukoswa GM, Charalambous S, Nelson G. The association between social capital and HIV treatment outcomes in south africa. \emph{PloS one}. 2017;\textbf{12}(11):e0184140.

10. Chola L, Alaba O. Association of neighbourhood and individual social capital, neighbourhood economic deprivation and self-rated health in south africa a multi-level analysis. \emph{PLoS ONE} {[}Internet{]}. 2013 Jul 29;\textbf{8}(7). Available from: \url{https://www.ncbi.nlm.nih.gov/pmc/articles/PMC3743525/}

11. Christian AK, Sanuade OA, Okyere MA, Adjaye-Gbewonyo K. Social capital is associated with improved subjective well-being of older adults with chronic non-communicable disease in six low- and middle-income countries. \emph{Globalization and Health} {[}Internet{]}. 2020 Jan 2;\textbf{16}. Available from: \url{https://www.ncbi.nlm.nih.gov/pmc/articles/PMC6941280/}

12. Cramm JM, Møller V, Nieboer AP. Individual- and neighbourhood-level indicators of subjective well-being in a small and poor eastern cape township: The effect of health, social capital, marital status, and income. \emph{Social Indicators Research} {[}Internet{]}. 2012;\textbf{105}(3):581--593. Available from: \url{https://www.ncbi.nlm.nih.gov/pmc/articles/PMC3249562/}

13. Lau YK, Ataguba JE. Investigating the relationship between self-rated health and social capital in south africa: A multilevel panel data analysis. \emph{BMC Public Health} {[}Internet{]}. 2015 Mar 19;\textbf{15}. Available from: \url{https://www.ncbi.nlm.nih.gov/pmc/articles/PMC4373512/}

14. Ramlagan S, Peltzer K, Phaswana-Mafuya N. Social capital and health among older adults in south africa. \emph{BMC Geriatrics} {[}Internet{]}. 2013 Sep 28;\textbf{13}:100. Available from: \url{https://www.ncbi.nlm.nih.gov/pmc/articles/PMC3851859/}

15. Adjaye-Gbewonyo D, Rebok GW, Gross AL, Gallo JJ, Underwood CR. Assessing urban-rural differences in the relationship between social capital and depression among ghanaian and south african older adults. \emph{PLoS ONE} {[}Internet{]}. 2019 Jun 19;\textbf{14}(6). Available from: \url{https://www.ncbi.nlm.nih.gov/pmc/articles/PMC6583957/}

16. Tomita A, Burns JK. A multilevel analysis of association between neighborhood social capital and depression: Evidence from the first south african national income dynamics study. \emph{Journal of affective disorders} {[}Internet{]}. 2013 Jan 10;\textbf{144}(1-2):101--105. Available from: \url{https://www.ncbi.nlm.nih.gov/pmc/articles/PMC3513630/}

17. Cramm JM, Koolman X, Møller V, Nieboer AP. Socio-economic status and self-reported tuberculosis: A multilevel analysis in a low-income township in the eastern cape, south africa. \emph{Journal of Public Health in Africa} {[}Internet{]}. 2011 Sep 5;\textbf{2}(2). Available from: \url{https://www.ncbi.nlm.nih.gov/pmc/articles/PMC5345507/}

18. Cramm JM, Koolman X, Møller V, Nieboer AP. Socio-economic status and self-reported tuberculosis: A multilevel analysis in a low-income township in the eastern cape, south africa. \emph{Journal of Public Health in Africa} {[}Internet{]}. 2011 Sep 5;\textbf{2}(2). Available from: \url{https://www.ncbi.nlm.nih.gov/pmc/articles/PMC5345507/}

19. Harling G, Kobayashi LC, Farrell MT, Wagner RG, Tollman S, Berkman L. Social contact, social support, and cognitive health in a population-based study of middle-aged and older men and women in rural south africa. \emph{Social Science \& Medicine (1982)} {[}Internet{]}. 2020 Sep;\textbf{260}. Available from: \url{https://www.ncbi.nlm.nih.gov/pmc/articles/PMC7441312/}

20. Siette J, Pomare C, Dodds L, Jorgensen M, Harrigan N, Georgiou A. \href{https://doi.org/10.1016/j.archger.2021.104525}{A comprehensive overview of social network measures for older adults: A systematic review}. \emph{Archives of Gerontology and Geriatrics}. 2021;\textbf{97}:104525.

21. Hernán M, Robins JM. Causal inference. Chapman \& Hall/CRC; 2018.

22. Halloran ME, Hudgens MG. Dependent happenings: A recent methodological review. \emph{Current epidemiology reports}. 2016;\textbf{3}(4):297--305.

23. Tchetgen Tchetgen EJ, Fulcher IR, Shpitser I. Auto-g-computation of causal effects on a network. \emph{Journal of the American Statistical Association}. 2020;112.

24. Statistics South Africa. General household survey 2019 {[}Internet{]}. Pretoria; 2021. Available from: \url{http://www.statssa.gov.za/publications/P0318/P03182019.pdf}

25. Statistics South Africa. Key statistics - bushbuckridge {[}Internet{]}. 2021. Available from: \url{http://www.statssa.gov.za/?page_id=993\&id=bushbuckridge-municipality}

26. Statistics South Africa. Key statistics - bushbuckridge {[}Internet{]}. 2021. Available from: \url{http://www.statssa.gov.za/?page_id=993\&id=bushbuckridge-municipality}

27. Statistics South Africa. General household survey 2019 {[}Internet{]}. Pretoria; 2021. Available from: \url{http://www.statssa.gov.za/publications/P0318/P03182019.pdf}

28. Kabudula CW, Houle B, Collinson MA, et al. Progression of the epidemiological transition in a rural south african setting: Findings from population surveillance in agincourt, 19932013. \emph{BMC Public Health} {[}Internet{]}. 2017 May 10;\textbf{17}. Available from: \url{https://www.ncbi.nlm.nih.gov/pmc/articles/PMC5424387/}

29. Payne CF, Gómez-Olivé FX, Kahn K, Berkman L. \href{https://doi.org/10.1093/geronb/gbx030}{Physical Function in an Aging Population in Rural South Africa: Findings From HAALSI and Cross-National Comparisons With HRS Sister Studies}. \emph{The Journals of Gerontology Series B, Psychological Sciences and Social Sciences}. 2017 Jul 1;\textbf{72}(4):665--679.

30. Rutstein SO, Johnson K, MEASURE OM, others. The DHS wealth index. ORC Macro, MEASURE DHS; 2004.

31. Kabudula CW, Houle B, Collinson MA, Kahn K, Tollman S, Clark S. Assessing Changes in Household Socioeconomic Status in Rural South Africa, 2001-2013: A Distributional Analysis Using Household Asset Indicators. \emph{Social Indicators Research} {[}Internet{]}. 2016 Jun 28;\textbf{133}(3):1047--1073. Available from: \url{http://dx.doi.org/10.1007/s11205-016-1397-z}

32. Besag J. Spatial interaction and the statistical analysis of lattice systems. \emph{Journal of the Royal Statistical Society: Series B (Methodological)}. 1974;\textbf{36}(2):192--225.

33. Buuren S van, Groothuis-Oudshoorn K. Mice: Multivariate imputation by chained equations in r. \emph{Journal of Statistical Software} {[}Internet{]}. 2011;\textbf{45}(3):1--67. Available from: \url{https://www.jstatsoft.org/v45/i03/}

34. Little RJA, Rubin DB. Statistical analysis with missing data. Third edition. Hoboken, NJ: Wiley; 2020.

35. Rubin DB. Inference and missing data. \emph{Biometrika} {[}Internet{]}. 1976;\textbf{63}(3):581--592. Available from: \url{https://academic.oup.com/biomet/article-lookup/doi/10.1093/biomet/63.3.581}

36. Haal K, Smith A, Doorslaer E van. \href{https://doi.org/10.1016/j.ssmph.2018.06.007}{The rise and fall of mortality inequality in South Africa in the HIV era}. \emph{SSM - population health}. 2018 Aug;\textbf{5}:239--248.

37. STATS SA. P0302: Mid-year population estimates 2022 {[}Internet{]}. 2022 Jul. Available from: \url{https://www.statssa.gov.za/publications/P0302/P03022022.pdf}

38. Kelly G. Disability, cash transfers and family practices in South Africa. \emph{Critical Social Policy} {[}Internet{]}. 2019 Aug 22;\textbf{39}(4):541--559. Available from: \url{http://dx.doi.org/10.1177/0261018319867593}

39. Moore E. Financing Social Reproduction. \emph{Revue internationale des études du développement} {[}Internet{]}. 2020;\textbf{N°242}(2):37. Available from: \url{http://dx.doi.org/10.3917/ried.242.0037}

40. Statistics South Africa. General household survey 2019 {[}Internet{]}. Pretoria; 2021. Available from: \url{http://www.statssa.gov.za/publications/P0318/P03182019.pdf}

41. Statistics South Africa. Key statistics - bushbuckridge {[}Internet{]}. 2021. Available from: \url{http://www.statssa.gov.za/?page_id=993\&id=bushbuckridge-municipality}

42. Makiwane M, Gumede NA, Makoae M, Vawda M. Family in a changing South Africa: structures, functions and the welfare of members. \emph{South African Review of Sociology} {[}Internet{]}. 2017 Apr 3;\textbf{48}(2):49--69. Available from: \url{https://www.tandfonline.com/doi/full/10.1080/21528586.2017.1288166}

43. Collinson MA, Mudzana T, Mutevedzi T, et al. Cohort Profile: South African Population Research Infrastructure Network (SAPRIN). \emph{International Journal of Epidemiology} {[}Internet{]}. 2021 Dec 30;\textbf{51}(4):e206--e216. Available from: \url{http://dx.doi.org/10.1093/ije/dyab261}

44. Collinson MA, White MJ, Ginsburg C, Gómez-Olivé FX, Kahn K, Tollman S. Youth migration, livelihood prospects and demographic dividend: A comparison of the census 2011 and agincourt health and demographic surveillance system in the rural northeast of south africa. \emph{African Population Studies} {[}Internet{]}. 2016 Sep 9; Available from: \url{http://dx.doi.org/10.11564/30-2-852}

45. Byass P, Sankoh O, Tollman SM, Högberg U, Wall S. Lessons from History for Designing and Validating Epidemiological Surveillance in Uncounted Populations. Althabe F, editor. \emph{PLoS ONE} {[}Internet{]}. 2011 Aug 3;\textbf{6}(8):e22897. Available from: \url{http://dx.doi.org/10.1371/journal.pone.0022897}

46. Vable AM, Gilsanz P, Kawachi I. \href{https://doi.org/10.1093/pubmed/fdz018}{Is it possible to overcome the 'long arm' of childhood socioeconomic disadvantage through upward socioeconomic mobility?} \emph{Journal of Public Health (Oxford, England)}. 2019 Sep 30;\textbf{41}(3):566--574.

47. Birnie K, Martin RM, Gallacher J, et al. \href{https://doi.org/10.1136/jech.2009.103648}{Socio-economic disadvantage from childhood to adulthood and locomotor function in old age: a lifecourse analysis of the Boyd Orr and Caerphilly prospective studies}. \emph{Journal of Epidemiology and Community Health}. 2011 Nov;\textbf{65}(11):1014--1023.

48. Landös A, Arx M von, Cheval B, et al. \href{https://doi.org/10.1093/eurpub/cky166}{Childhood socioeconomic circumstances and disability trajectories in older men and women: a European cohort study}. \emph{European Journal of Public Health}. 2019 Feb 1;\textbf{29}(1):50--58.

49. Kelly G. The influence of disability-related cash transfers on family practices in south africa. \emph{CSSR Working Paper No}. 2018;\textbf{414}:2006.

50. Harper S, Seekings J. Claims on and obligations to kin in cape town, south africa. 2010;

51. Madhavan S, Clark S, Beguy D, Kabiru CW, Gross M. \href{https://doi.org/10.1080/00324728.2016.1262965}{Moving beyond the household: Innovations in data collection on kinship}. \emph{Population Studies}. 2017 Mar;\textbf{71}(1):117--132.

52. Shalizi CR, Thomas AC. Homophily and contagion are generically confounded in observational social network studies. \emph{Sociological methods \& research}. 2011;\textbf{40}(2):211--239.

\clearpage
\appendix
\begin{center}
{\Large\bfseries Supplementary Materials}
\end{center}

\hypertarget{supp-appendix-s1-definition-and-identification-of-causal-estimands}{%
\section{Appendix S1: Definition and identification of causal estimands}\label{supp-appendix-s1-definition-and-identification-of-causal-estimands}}

\hypertarget{supp-notation}{%
\subsection{Notation}\label{supp-notation}}

Unless otherwise indicated, unbolded Latin letters are used to indicate scalar quantities and bold letters indicate vectors. For vector \(\mathbf{V}\) and set \(A\), \(\mathbf{V}_{A}\) is the sub-vector of \(\mathbf{V}\) consisting of entries whose indices are included in \(A\). Random variables are indicated using capital letters and the realization of a given random variable is indicated using a small letter. (e.g.~if \(R\) is a random variable, \(r\) is a particular value of \(R\)). Sets are indicated using unbolded capital letters. The set union operator is \(\cup\) and the intersection operator is \(\cap\). For sets \(A\) and \(B\), \(A\backslash B\) is a set containing all the elements of \(A\) except for those contained by \(B\).

\hypertarget{supp-definition-and-identification-strategy}{%
\subsection{Definition and identification strategy}\label{supp-definition-and-identification-strategy}}

We define a \(n \times 1\) vector of dichotomous potential outcomes \(\mathbf{Y}\left( \mathbf{a} \right)\) under \(n \times 1\) treatment vector \(\mathbf{a}\). The \(i^{th}\) element of this vector, \(Y_{i}\left( \mathbf{a} \right)\), is the potential outcome for household \(i\) under the level of household wealth dictated by \(\mathbf{a}\). The treatment for household \(i\) is \(a_{i}\) - the \(i^{th}\) element of \(\mathbf{a}\). \(\mathbf{L}\), \(\mathbf{Y}\), and \(\mathbf{A}\) are \(n\)-dimensional vectors of covariates, observed outcomes, and observed treatment, respectively.

We assume exchangeability and consistency:

\[\mathbf{Y}\left( \mathbf{a} \right)\bot\mathbf{A}|\mathbf{L}\text{ and }\mathbf{Y} = \mathbf{Y}\left( \mathbf{A} \right)\text{.}\]

These assumptions imply that:

\[E\left\lbrack \mathbf{Y}\left( \mathbf{a} \right)|\mathbf{L} \right\rbrack = E\left\lbrack \mathbf{Y}\left( \mathbf{a} \right)|\mathbf{L},\mathbf{A} = \mathbf{a} \right\rbrack = E\left\lbrack \mathbf{Y}|\mathbf{L},\mathbf{A} = \mathbf{a} \right\rbrack\text{.}\]

We define \(n \times n\) matrix \(\mathbf{D}\):

\[\begin{aligned}
\mathbf{D} &= \frac{d}{d\mathbf{a}}E\left\lbrack \mathbf{Y}\left( \mathbf{a} \right)|\mathbf{L} \right\rbrack \\
 &= \frac{d}{d\mathbf{a}}E\left\lbrack \mathbf{Y}|\mathbf{L,A = a} \right\rbrack
\end{aligned}\]

Then the causal estimands are defined:

\[\begin{bmatrix}
ATE \\
ADE \\
ASE
\end{bmatrix} = \frac{1}{n}\begin{bmatrix}
\text{grandsum}\left( \mathbf{D} \right) \\
\text{trace}\left( \mathbf{D} \right) \\
\text{grandsum}\left( \mathbf{D} \right) - \text{trace}\left( \mathbf{D} \right)
\end{bmatrix}\]

where \(\text{grandsum}\left( \mathbf{D} \right)\) is the sum of the entries of matrix \(\mathbf{D}\).

To estimate \(\mathbf{D}\) from observed data, it is necessary to specify restrictions on the joint distribution of vector-valued random variable \(\left( \mathbf{Y},\mathbf{L},\mathbf{A} \right)\). (Note that the entire data set represents one realization of this vector of random variables).

To this end, we assume that observed data arise from a Markov random field defined by local Markov condition, which formally stated:

\[Y_{i}\bot Y_{j}|\mathbf{L}_{\mathcal{N}_{i}},\mathbf{A}_{\mathcal{N}_{i}},\mathbf{Y}_{\mathcal{N}_{i}\backslash\{ i\}}\text{ where }j \notin \mathcal{N}_{i}\]

where set \(\mathcal{N}_{i}\) contains the the index for household \(i\) as well as the indices of households connected to household \(i\) in the network. \(\mathbf{L}_{\mathcal{N}_{i}}\) is the sub-vector of \(\mathbf{L}\) consisting of elements whose indices are included in set \(\mathcal{N}_{i}\). \(\mathbf{A}_{\mathcal{N}_{i}}\) and \(\mathbf{Y}_{\mathcal{N}_{i}\backslash\{ i\}}\) are similarly defined.

We further assume a functional form for the conditional mean of \(\mathbf{Y}\):

\[\begin{aligned}
E\left\lbrack \mathbf{Y}|\mathbf{L},\mathbf{A},\mathbf{Y} \right\rbrack = {} & \beta_{0}\mathbf{1} + \beta_{1}\mathbf{A} + \beta_{2}\mathbf{L} + \beta_{3}\mathbf{MA} + \beta_{4}\mathbf{ML} \\
 & + \beta_{5}\mathbf{A} \odot \mathbf{L} + \beta_{6}\mathbf{L} \odot \mathbf{MA} + \beta_{7}\mathbf{L} \odot \mathbf{ML} + \phi\mathbf{MY}
\end{aligned}\]

where \(\mathbf{M}\) is the adjacency matrix representing the network, \(\mathbf{1}\) is a \(n \times 1\) vector of \(1\)'s, and \(\odot\) is the element-wise matrix multiplication operator. (Note that \(\mathbf{M}\), by definition, has zeros on the main diagonal).

By taking iterative expectations, gathering terms involving \(E\left\lbrack \mathbf{Y}|\mathbf{L},\mathbf{A} = \mathbf{a} \right\rbrack\), and taking the first derivative with respect to vector \(\mathbf{a}\), we find a closed-form function for \(\mathbf{D}\), yielding a formula for the causal estimands:

\[\begin{aligned}
\mathbf{D} &= \frac{d}{d\mathbf{a}}E\left\lbrack \mathbf{Y}|\mathbf{L},\mathbf{A} = \mathbf{a} \right\rbrack \\
 &= \left( \mathbf{I} - \phi\mathbf{M} \right)^{- 1}\left( \beta_{1}\mathbf{I} + \beta_{3}\mathbf{M} + \beta_{5}\text{diag}\left( \mathbf{L} \right) + \beta_{6}\text{diag}\left( \mathbf{L} \right) \right)
\end{aligned}\]

where \(\text{diag}\left( \mathbf{L} \right)\) is the \(n \times n\) matrix with entries of \(n\)-dimensional vector \(\mathbf{L}\) on the main diagonal and zeros elsewhere.

The parameter \(\phi\) represents the degree of network outcome dependence between deaths among connected households beyond that induced by \(\left( \mathbf{L}_{t},\mathbf{A}_{t} \right)\), and must be incorporated to account for such dependence in identifying total, direct, and spillover causal effects on a network of interconnected units.\textsuperscript{23} In the special case where \(\phi = 0\), the causal quantities of interest are given by:

\[\begin{aligned}
ASE &= \beta_{3}\bar{d} + \beta_{6}\frac{1}{n}\sum_{i}^{}{l_{i}d_{i}} \\
ADE &= \beta_{1} + \beta_{5}\bar{l}
\end{aligned}\]

where \(d_{i}\) is the number of households connected to household \(i\) and \(\bar{d}\) is the average of \(d_{i}\) across households.

\hypertarget{supp-appendix-s2-missing-data}{%
\section{Appendix S2: Missing Data}\label{supp-appendix-s2-missing-data}}

To account for missing data, we implemented multiple imputation using a custom-built R-package based on the Multiple Imputation by Chained Equations (MICE) package. Household wealth was imputed using predictive mean matching. Network wealth was computed using imputed household wealth values. We created 8 complete datasets, each time iterating the custom-built MICE algorithm 60 times. Robust standard errors were calculated for each imputation. Coefficient estimates and standard errors were pooled across imputations using Rubin's rules. The table shown below lists all variables included as predictors in the imputation model. The imputation model produces valid inferences under the assumption that data were ``missing at random'' and that the functional forms of regressions used for imputation are correct.

{\footnotesize\setlength{\tabcolsep}{2.5pt}
\begin{longtable}[]{@{}
  >{\raggedright\arraybackslash}p{(\columnwidth - 6\tabcolsep) * \real{0.1800}}
  >{\raggedright\arraybackslash}p{(\columnwidth - 6\tabcolsep) * \real{0.1862}}
  >{\raggedright\arraybackslash}p{(\columnwidth - 6\tabcolsep) * \real{0.3188}}
  >{\raggedright\arraybackslash}p{(\columnwidth - 6\tabcolsep) * \real{0.3150}}@{}}
\caption*{Table S1: Variables included in the multiple imputation model}\tabularnewline
\toprule\noalign{}
\textbf{Imputed} & \textbf{Computed} & \multicolumn{2}{>{\raggedright\arraybackslash}p{(\columnwidth - 6\tabcolsep) * \real{0.6338} + 2\tabcolsep}@{}}{%
\textbf{Complete}} \\
\midrule\noalign{}
\endfirsthead
\toprule\noalign{}
\textbf{Imputed} & \textbf{Computed} & \multicolumn{2}{>{\raggedright\arraybackslash}p{(\columnwidth - 6\tabcolsep) * \real{0.6338} + 2\tabcolsep}@{}}{%
\textbf{Complete}} \\
\midrule\noalign{}
\endhead
\bottomrule\noalign{}
\endlastfoot
Household Wealth$_{t}$ & Network Wealth$_{t}$

Network Wealth$_{t - 1}$

Household Wealth$_{t - 1}$ & Count: Connected Households$_{t}$

Count: Children$_{t}$

Count: Adult Men$_{t}$

Count: Elder Men$_{t}$

Count: Adult Women$_{t}$

Count: Elder Women$_{t}$

Deaths: Children$_{t}$

Deaths: Adult Men$_{t}$

Deaths: Elder Men$_{t}$

Deaths: Adult Women$_{t}$

Deaths: Elder Women$_{t}$

Count: Currently Working$_{t}$

Count: Not Working$_{t}$

Count: Unemployed$_{t}$

Count: Not in labor market$_{t}$ & \(Count:\ Connected\ {Households}_{t - 1}\)

Count: Adult Men$_{t - 1}$

Count: Elder Men$_{t - 1}$

Count: Adult Women$_{t - 1}$

Count: Elder Women$_{t - 1}$

Deaths: Children$_{t - 1}$

Deaths: Adult Men$_{t - 1}$

Deaths: Elder Men$_{t - 1}$

Deaths: Adult Women$_{t - 1}$

Deaths: Elder Women$_{t - 1}$

Count: Currently Working$_{t - 1}$

Count: Not Working$_{t - 1}$

Count: Unemployed$_{t - 1}$

Count: Not in labor market$_{t - 1}$ \\
\end{longtable}
}

\hypertarget{supp-appendix-s3-descriptive-statistics}{%
\section{Appendix S3: Descriptive Statistics}\label{supp-appendix-s3-descriptive-statistics}}

\hypertarget{supp-original-villages-demographic-variables}{%
\subsection{Original Villages Demographic Variables}\label{supp-original-villages-demographic-variables}}

{\footnotesize\setlength{\tabcolsep}{2.5pt}
\begin{longtable}[]{@{}lrrrrr@{}}
\toprule\noalign{}
 & 2009 & 2010 & 2011 & 2012 & 2013 \\
\midrule\noalign{}
\endhead
\bottomrule\noalign{}
\endlastfoot
\multicolumn{6}{@{}l}{\textbf{Number of Households}} \\
N & 16275 (100.0\%) & 16629 (100.0\%) & 17047 (100.0\%) & 17539 (100.0\%) & 17836 (100.0\%) \\
\multicolumn{6}{@{}l}{\textbf{Village Type}} \\
Original Villages & 16275 (100.0\%) & 16629 (100.0\%) & 17047 (100.0\%) & 17539 (100.0\%) & 17836 (100.0\%) \\
\multicolumn{6}{@{}l}{\textbf{\# Connected Households}} \\
0 & 4769 (29.3\%) & 4807 (28.9\%) & 4884 (28.7\%) & 5097 (29.1\%) & 5125 (28.7\%) \\
1 & 3372 (20.7\%) & 3381 (20.3\%) & 3455 (20.3\%) & 3521 (20.1\%) & 3561 (20.0\%) \\
2 & 2734 (16.8\%) & 2832 (17.0\%) & 2917 (17.1\%) & 2908 (16.6\%) & 2964 (16.6\%) \\
3 & 1943 (11.9\%) & 2038 (12.3\%) & 2066 (12.1\%) & 2164 (12.3\%) & 2177 (12.2\%) \\
4 & 1332 (8.2\%) & 1361 (8.2\%) & 1416 (8.3\%) & 1441 (8.2\%) & 1484 (8.3\%) \\
5 & 799 (4.9\%) & 851 (5.1\%) & 906 (5.3\%) & 928 (5.3\%) & 962 (5.4\%) \\
6+ & 1326 (8.1\%) & 1359 (8.2\%) & 1403 (8.2\%) & 1480 (8.4\%) & 1563 (8.8\%) \\
\multicolumn{6}{@{}l}{\textbf{\# Household Members}} \\
1 & 1647 (10.1\%) & 1697 (10.2\%) & 1788 (10.5\%) & 1952 (11.1\%) & 1987 (11.1\%) \\
2-5 & 7380 (45.3\%) & 7655 (46.0\%) & 7958 (46.7\%) & 8162 (46.5\%) & 8372 (46.9\%) \\
6-10 & 5751 (35.3\%) & 5790 (34.8\%) & 5805 (34.1\%) & 5933 (33.8\%) & 5981 (33.5\%) \\
11-20 & 1458 (9.0\%) & 1442 (8.7\%) & 1461 (8.6\%) & 1461 (8.3\%) & 1463 (8.2\%) \\
21+ & 39 (0.2\%) & 45 (0.3\%) & 35 (0.2\%) & 31 (0.2\%) & 33 (0.2\%) \\
\multicolumn{6}{@{}l}{\textbf{\# Children (\textless5)}} \\
0 & 8234 (50.6\%) & 8596 (51.7\%) & 8952 (52.5\%) & 9352 (53.3\%) & 9671 (54.2\%) \\
1 & 5415 (33.3\%) & 5485 (33.0\%) & 5564 (32.6\%) & 5602 (31.9\%) & 5578 (31.3\%) \\
2-5 & 2607 (16.0\%) & 2531 (15.2\%) & 2515 (14.8\%) & 2572 (14.7\%) & 2579 (14.5\%) \\
6-10 & 19 (0.1\%) & 17 (0.1\%) & 16 (0.1\%) & 13 (0.1\%) & 7 (0.0\%) \\
11-20 & 0 (0.0\%) & 0 (0.0\%) & 0 (0.0\%) & 0 (0.0\%) & 1 (0.0\%) \\
\multicolumn{6}{@{}l}{\textbf{\# Adults (15-59)}} \\
0 & 606 (3.7\%) & 620 (3.7\%) & 653 (3.8\%) & 684 (3.9\%) & 715 (4.0\%) \\
1 & 2490 (15.3\%) & 2564 (15.4\%) & 2669 (15.7\%) & 2824 (16.1\%) & 2871 (16.1\%) \\
2-5 & 10695 (65.7\%) & 10945 (65.8\%) & 11150 (65.4\%) & 11444 (65.2\%) & 11657 (65.4\%) \\
6-10 & 2381 (14.6\%) & 2402 (14.4\%) & 2469 (14.5\%) & 2488 (14.2\%) & 2486 (13.9\%) \\
11-20 & 101 (0.6\%) & 96 (0.6\%) & 104 (0.6\%) & 97 (0.6\%) & 106 (0.6\%) \\
21+ & 2 (0.0\%) & 2 (0.0\%) & 2 (0.0\%) & 2 (0.0\%) & 1 (0.0\%) \\
\multicolumn{6}{@{}l}{\textbf{\# Elders (60+)}} \\
0 & 11510 (70.7\%) & 11699 (70.4\%) & 12044 (70.7\%) & 12348 (70.4\%) & 12506 (70.1\%) \\
1 & 3797 (23.3\%) & 3923 (23.6\%) & 4018 (23.6\%) & 4165 (23.7\%) & 4287 (24.0\%) \\
2 & 914 (5.6\%) & 954 (5.7\%) & 937 (5.5\%) & 971 (5.5\%) & 992 (5.6\%) \\
3+ & 54 (0.3\%) & 53 (0.3\%) & 48 (0.3\%) & 55 (0.3\%) & 51 (0.3\%) \\
\multicolumn{6}{@{}l}{\textbf{\# Adult Men}} \\
0 & 2832 (17.4\%) & 2949 (17.7\%) & 3040 (17.8\%) & 3160 (18.0\%) & 3183 (17.8\%) \\
1 & 6544 (40.2\%) & 6687 (40.2\%) & 6908 (40.5\%) & 7112 (40.5\%) & 7290 (40.9\%) \\
2 & 3502 (21.5\%) & 3549 (21.3\%) & 3602 (21.1\%) & 3738 (21.3\%) & 3783 (21.2\%) \\
3 & 1947 (12.0\%) & 1974 (11.9\%) & 2013 (11.8\%) & 2048 (11.7\%) & 2093 (11.7\%) \\
4 & 858 (5.3\%) & 904 (5.4\%) & 939 (5.5\%) & 926 (5.3\%) & 941 (5.3\%) \\
5 & 390 (2.4\%) & 380 (2.3\%) & 348 (2.0\%) & 356 (2.0\%) & 349 (2.0\%) \\
6+ & 202 (1.2\%) & 186 (1.1\%) & 197 (1.2\%) & 199 (1.1\%) & 197 (1.1\%) \\
\multicolumn{6}{@{}l}{\textbf{\# Adult Women}} \\
0 & 2408 (14.8\%) & 2484 (14.9\%) & 2615 (15.3\%) & 2799 (16.0\%) & 2908 (16.3\%) \\
1 & 6065 (37.3\%) & 6213 (37.4\%) & 6375 (37.4\%) & 6549 (37.3\%) & 6613 (37.1\%) \\
2 & 3964 (24.4\%) & 4073 (24.5\%) & 4073 (23.9\%) & 4093 (23.3\%) & 4168 (23.4\%) \\
3 & 2189 (13.5\%) & 2202 (13.2\%) & 2285 (13.4\%) & 2358 (13.4\%) & 2375 (13.3\%) \\
4 & 1019 (6.3\%) & 1031 (6.2\%) & 1043 (6.1\%) & 1068 (6.1\%) & 1100 (6.2\%) \\
5 & 408 (2.5\%) & 399 (2.4\%) & 437 (2.6\%) & 435 (2.5\%) & 435 (2.4\%) \\
6+ & 222 (1.4\%) & 227 (1.4\%) & 219 (1.3\%) & 237 (1.4\%) & 237 (1.3\%) \\
\multicolumn{6}{@{}l}{\textbf{\# Elder Men}} \\
0 & 14298 (87.9\%) & 14542 (87.4\%) & 14936 (87.6\%) & 15313 (87.3\%) & 15566 (87.3\%) \\
1 & 1959 (12.0\%) & 2074 (12.5\%) & 2094 (12.3\%) & 2209 (12.6\%) & 2254 (12.6\%) \\
2 & 18 (0.1\%) & 12 (0.1\%) & 16 (0.1\%) & 16 (0.1\%) & 16 (0.1\%) \\
3 & 0 (0.0\%) & 1 (0.0\%) & 1 (0.0\%) & 1 (0.0\%) & 0 (0.0\%) \\
\multicolumn{6}{@{}l}{\textbf{\# Elder Women}} \\
0 & 12608 (77.5\%) & 12871 (77.4\%) & 13263 (77.8\%) & 13642 (77.8\%) & 13829 (77.5\%) \\
1 & 3545 (21.8\%) & 3628 (21.8\%) & 3662 (21.5\%) & 3766 (21.5\%) & 3878 (21.7\%) \\
2 & 119 (0.7\%) & 127 (0.8\%) & 120 (0.7\%) & 130 (0.7\%) & 127 (0.7\%) \\
3 & 3 (0.0\%) & 3 (0.0\%) & 2 (0.0\%) & 1 (0.0\%) & 2 (0.0\%) \\
\multicolumn{6}{@{}l}{\textbf{\# Deaths in Household}} \\
0 & 15394 (94.6\%) & 15835 (95.2\%) & 16294 (95.6\%) & 16808 (95.8\%) & 17127 (96.0\%) \\
1 & 851 (5.2\%) & 767 (4.6\%) & 722 (4.2\%) & 710 (4.0\%) & 690 (3.9\%) \\
2 & 30 (0.2\%) & 27 (0.2\%) & 30 (0.2\%) & 20 (0.1\%) & 19 (0.1\%) \\
3+ & 0 (0.0\%) & 0 (0.0\%) & 1 (0.0\%) & 1 (0.0\%) & 0 (0.0\%) \\
\end{longtable}
}

{\footnotesize\setlength{\tabcolsep}{2.5pt}
\begin{longtable}[]{@{}lrrrrr@{}}
\toprule\noalign{}
 & 2014 & 2015 & 2016 & 2017 & 2018 \\
\midrule\noalign{}
\endhead
\bottomrule\noalign{}
\endlastfoot
\multicolumn{6}{@{}l}{\textbf{Number of Households}} \\
N & 18144 (100.0\%) & 18332 (100.0\%) & 18461 (100.0\%) & 18337 (100.0\%) & 17924 (100.0\%) \\
\multicolumn{6}{@{}l}{\textbf{Village Type}} \\
Original Villages & 18144 (100.0\%) & 18332 (100.0\%) & 18461 (100.0\%) & 18337 (100.0\%) & 17924 (100.0\%) \\
\multicolumn{6}{@{}l}{\textbf{\# Connected Households}} \\
0 & 5165 (28.5\%) & 5168 (28.2\%) & 5138 (27.8\%) & 5084 (27.7\%) & 5107 (28.5\%) \\
1 & 3589 (19.8\%) & 3618 (19.7\%) & 3633 (19.7\%) & 3722 (20.3\%) & 3776 (21.1\%) \\
2 & 2981 (16.4\%) & 3025 (16.5\%) & 3117 (16.9\%) & 3102 (16.9\%) & 3022 (16.9\%) \\
3 & 2215 (12.2\%) & 2249 (12.3\%) & 2315 (12.5\%) & 2287 (12.5\%) & 2212 (12.3\%) \\
4 & 1554 (8.6\%) & 1618 (8.8\%) & 1581 (8.6\%) & 1559 (8.5\%) & 1474 (8.2\%) \\
5 & 1036 (5.7\%) & 1010 (5.5\%) & 1027 (5.6\%) & 1000 (5.5\%) & 942 (5.3\%) \\
6+ & 1604 (8.8\%) & 1644 (9.0\%) & 1650 (8.9\%) & 1583 (8.6\%) & 1391 (7.8\%) \\
\multicolumn{6}{@{}l}{\textbf{\# Household Members}} \\
1 & 2022 (11.1\%) & 2040 (11.1\%) & 2134 (11.6\%) & 2181 (11.9\%) & 2148 (12.0\%) \\
2-5 & 8555 (47.2\%) & 8701 (47.5\%) & 8786 (47.6\%) & 8711 (47.5\%) & 8500 (47.4\%) \\
6-10 & 6042 (33.3\%) & 6086 (33.2\%) & 6047 (32.8\%) & 6000 (32.7\%) & 5909 (33.0\%) \\
11-20 & 1503 (8.3\%) & 1480 (8.1\%) & 1473 (8.0\%) & 1420 (7.7\%) & 1345 (7.5\%) \\
21+ & 22 (0.1\%) & 25 (0.1\%) & 21 (0.1\%) & 25 (0.1\%) & 22 (0.1\%) \\
\multicolumn{6}{@{}l}{\textbf{\# Children (\textless5)}} \\
0 & 9905 (54.6\%) & 10041 (54.8\%) & 10377 (56.2\%) & 10652 (58.1\%) & 11073 (61.8\%) \\
1 & 5671 (31.3\%) & 5745 (31.3\%) & 5722 (31.0\%) & 5515 (30.1\%) & 5151 (28.7\%) \\
2-5 & 2558 (14.1\%) & 2536 (13.8\%) & 2355 (12.8\%) & 2166 (11.8\%) & 1698 (9.5\%) \\
6-10 & 10 (0.1\%) & 10 (0.1\%) & 7 (0.0\%) & 4 (0.0\%) & 2 (0.0\%) \\
\multicolumn{6}{@{}l}{\textbf{\# Adults (15-59)}} \\
0 & 710 (3.9\%) & 711 (3.9\%) & 737 (4.0\%) & 738 (4.0\%) & 729 (4.1\%) \\
1 & 2903 (16.0\%) & 2897 (15.8\%) & 2930 (15.9\%) & 2902 (15.8\%) & 2780 (15.5\%) \\
2-5 & 11931 (65.8\%) & 12107 (66.0\%) & 12161 (65.9\%) & 12136 (66.2\%) & 11903 (66.4\%) \\
6-10 & 2494 (13.7\%) & 2505 (13.7\%) & 2525 (13.7\%) & 2461 (13.4\%) & 2424 (13.5\%) \\
11-20 & 105 (0.6\%) & 110 (0.6\%) & 106 (0.6\%) & 98 (0.5\%) & 86 (0.5\%) \\
21+ & 1 (0.0\%) & 2 (0.0\%) & 2 (0.0\%) & 2 (0.0\%) & 2 (0.0\%) \\
\multicolumn{6}{@{}l}{\textbf{\# Elders (60+)}} \\
0 & 12693 (70.0\%) & 12810 (69.9\%) & 12819 (69.4\%) & 12600 (68.7\%) & 12062 (67.3\%) \\
1 & 4372 (24.1\%) & 4436 (24.2\%) & 4549 (24.6\%) & 4645 (25.3\%) & 4743 (26.5\%) \\
2 & 1029 (5.7\%) & 1036 (5.7\%) & 1048 (5.7\%) & 1049 (5.7\%) & 1076 (6.0\%) \\
3+ & 50 (0.3\%) & 50 (0.3\%) & 45 (0.2\%) & 43 (0.2\%) & 43 (0.2\%) \\
\multicolumn{6}{@{}l}{\textbf{\# Adult Men}} \\
0 & 3197 (17.6\%) & 3212 (17.5\%) & 3222 (17.5\%) & 3151 (17.2\%) & 3015 (16.8\%) \\
1 & 7439 (41.0\%) & 7537 (41.1\%) & 7612 (41.2\%) & 7543 (41.1\%) & 7257 (40.5\%) \\
2 & 3880 (21.4\%) & 3902 (21.3\%) & 3887 (21.1\%) & 3922 (21.4\%) & 3959 (22.1\%) \\
3 & 2092 (11.5\%) & 2110 (11.5\%) & 2169 (11.7\%) & 2136 (11.6\%) & 2132 (11.9\%) \\
4 & 976 (5.4\%) & 995 (5.4\%) & 1001 (5.4\%) & 1028 (5.6\%) & 1005 (5.6\%) \\
5 & 378 (2.1\%) & 393 (2.1\%) & 399 (2.2\%) & 372 (2.0\%) & 364 (2.0\%) \\
6+ & 182 (1.0\%) & 183 (1.0\%) & 171 (0.9\%) & 185 (1.0\%) & 192 (1.1\%) \\
\multicolumn{6}{@{}l}{\textbf{\# Adult Women}} \\
0 & 2979 (16.4\%) & 3034 (16.6\%) & 3166 (17.1\%) & 3251 (17.7\%) & 3257 (18.2\%) \\
1 & 6702 (36.9\%) & 6718 (36.6\%) & 6664 (36.1\%) & 6455 (35.2\%) & 6170 (34.4\%) \\
2 & 4298 (23.7\%) & 4335 (23.6\%) & 4340 (23.5\%) & 4426 (24.1\%) & 4380 (24.4\%) \\
3 & 2392 (13.2\%) & 2457 (13.4\%) & 2507 (13.6\%) & 2456 (13.4\%) & 2403 (13.4\%) \\
4 & 1116 (6.2\%) & 1117 (6.1\%) & 1138 (6.2\%) & 1134 (6.2\%) & 1090 (6.1\%) \\
5 & 426 (2.3\%) & 439 (2.4\%) & 420 (2.3\%) & 386 (2.1\%) & 407 (2.3\%) \\
6+ & 231 (1.3\%) & 232 (1.3\%) & 226 (1.2\%) & 229 (1.2\%) & 217 (1.2\%) \\
\multicolumn{6}{@{}l}{\textbf{\# Elder Men}} \\
0 & 15835 (87.3\%) & 15989 (87.2\%) & 16080 (87.1\%) & 15905 (86.7\%) & 15416 (86.0\%) \\
1 & 2289 (12.6\%) & 2326 (12.7\%) & 2365 (12.8\%) & 2422 (13.2\%) & 2495 (13.9\%) \\
2 & 20 (0.1\%) & 17 (0.1\%) & 16 (0.1\%) & 10 (0.1\%) & 13 (0.1\%) \\
\multicolumn{6}{@{}l}{\textbf{\# Elder Women}} \\
0 & 14029 (77.3\%) & 14168 (77.3\%) & 14202 (76.9\%) & 14024 (76.5\%) & 13539 (75.5\%) \\
1 & 3982 (21.9\%) & 4034 (22.0\%) & 4137 (22.4\%) & 4198 (22.9\%) & 4268 (23.8\%) \\
2 & 130 (0.7\%) & 125 (0.7\%) & 119 (0.6\%) & 111 (0.6\%) & 115 (0.6\%) \\
3 & 3 (0.0\%) & 5 (0.0\%) & 3 (0.0\%) & 4 (0.0\%) & 2 (0.0\%) \\
\multicolumn{6}{@{}l}{\textbf{\# Deaths in Household}} \\
0 & 17445 (96.1\%) & 17629 (96.2\%) & 17744 (96.1\%) & 17737 (96.7\%) & 17470 (97.5\%) \\
1 & 677 (3.7\%) & 684 (3.7\%) & 697 (3.8\%) & 584 (3.2\%) & 450 (2.5\%) \\
2 & 21 (0.1\%) & 19 (0.1\%) & 20 (0.1\%) & 16 (0.1\%) & 4 (0.0\%) \\
3+ & 1 (0.0\%) & 0 (0.0\%) & 0 (0.0\%) & 0 (0.0\%) & 0 (0.0\%) \\
\end{longtable}
}

\hypertarget{supp-original-villages-economic-variables}{%
\subsection{Original Villages Economic Variables}\label{supp-original-villages-economic-variables}}

{\footnotesize\setlength{\tabcolsep}{2.5pt}
\begin{longtable}[]{@{}lrrrrr@{}}
\toprule\noalign{}
 & 2009 & 2010 & 2011 & 2012 & 2013 \\
\midrule\noalign{}
\endhead
\bottomrule\noalign{}
\endlastfoot
\multicolumn{6}{@{}l}{\textbf{Number of Households}} \\
N & 16275 (100.0\%) & 16629 (100.0\%) & 17047 (100.0\%) & 17539 (100.0\%) & 17836 (100.0\%) \\
\multicolumn{6}{@{}l}{\textbf{Village Type}} \\
Original Villages & 16275 (100.0\%) & 16629 (100.0\%) & 17047 (100.0\%) & 17539 (100.0\%) & 17836 (100.0\%) \\
\multicolumn{6}{@{}l}{\textbf{\# Not in Employment Market}} \\
0 & 4180 (25.7\%) & 4641 (27.9\%) & 5091 (29.9\%) & 5567 (31.7\%) & 17832 (100.0\%) \\
1 & 3770 (23.2\%) & 3878 (23.3\%) & 4055 (23.8\%) & 4231 (24.1\%) & 4 (0.0\%) \\
2 & 3162 (19.4\%) & 3237 (19.5\%) & 3242 (19.0\%) & 3291 (18.8\%) & 0 (0.0\%) \\
3 & 2433 (14.9\%) & 2331 (14.0\%) & 2291 (13.4\%) & 2231 (12.7\%) & 0 (0.0\%) \\
4 & 1399 (8.6\%) & 1335 (8.0\%) & 1253 (7.4\%) & 1185 (6.8\%) & 0 (0.0\%) \\
5 & 707 (4.3\%) & 659 (4.0\%) & 628 (3.7\%) & 609 (3.5\%) & 0 (0.0\%) \\
6+ & 624 (3.8\%) & 548 (3.3\%) & 487 (2.9\%) & 425 (2.4\%) & 0 (0.0\%) \\
\multicolumn{6}{@{}l}{\textbf{\# Not Working}} \\
0 & 2712 (16.7\%) & 3055 (18.4\%) & 3392 (19.9\%) & 3751 (21.4\%) & 17825 (99.9\%) \\
1 & 3443 (21.2\%) & 3661 (22.0\%) & 3952 (23.2\%) & 4247 (24.2\%) & 11 (0.1\%) \\
2 & 3013 (18.5\%) & 3138 (18.9\%) & 3181 (18.7\%) & 3259 (18.6\%) & 0 (0.0\%) \\
3 & 2541 (15.6\%) & 2540 (15.3\%) & 2565 (15.0\%) & 2544 (14.5\%) & 0 (0.0\%) \\
4 & 1894 (11.6\%) & 1827 (11.0\%) & 1760 (10.3\%) & 1722 (9.8\%) & 0 (0.0\%) \\
5 & 1260 (7.7\%) & 1154 (6.9\%) & 1069 (6.3\%) & 984 (5.6\%) & 0 (0.0\%) \\
6+ & 1412 (8.7\%) & 1254 (7.5\%) & 1128 (6.6\%) & 1032 (5.9\%) & 0 (0.0\%) \\
\multicolumn{6}{@{}l}{\textbf{\# Currently Working}} \\
0 & 4936 (30.3\%) & 5380 (32.4\%) & 5858 (34.4\%) & 6371 (36.3\%) & 17828 (100.0\%) \\
1 & 6408 (39.4\%) & 6592 (39.6\%) & 6800 (39.9\%) & 7018 (40.0\%) & 8 (0.0\%) \\
2 & 3155 (19.4\%) & 3060 (18.4\%) & 2967 (17.4\%) & 2868 (16.4\%) & 0 (0.0\%) \\
3 & 1143 (7.0\%) & 1068 (6.4\%) & 974 (5.7\%) & 885 (5.0\%) & 0 (0.0\%) \\
4 & 449 (2.8\%) & 370 (2.2\%) & 322 (1.9\%) & 288 (1.6\%) & 0 (0.0\%) \\
5 & 117 (0.7\%) & 108 (0.6\%) & 84 (0.5\%) & 72 (0.4\%) & 0 (0.0\%) \\
6+ & 67 (0.4\%) & 51 (0.3\%) & 42 (0.2\%) & 37 (0.2\%) & 0 (0.0\%) \\
\multicolumn{6}{@{}l}{\textbf{\# Unemployed}} \\
0 & 2924 (18.0\%) & 3236 (19.5\%) & 3602 (21.1\%) & 3974 (22.7\%) & 17821 (99.9\%) \\
1 & 4824 (29.6\%) & 5194 (31.2\%) & 5558 (32.6\%) & 5978 (34.1\%) & 15 (0.1\%) \\
2 & 4503 (27.7\%) & 4537 (27.3\%) & 4537 (26.6\%) & 4492 (25.6\%) & 0 (0.0\%) \\
3 & 2121 (13.0\%) & 2004 (12.1\%) & 1878 (11.0\%) & 1778 (10.1\%) & 0 (0.0\%) \\
4 & 1032 (6.3\%) & 930 (5.6\%) & 854 (5.0\%) & 780 (4.4\%) & 0 (0.0\%) \\
5 & 518 (3.2\%) & 448 (2.7\%) & 402 (2.4\%) & 348 (2.0\%) & 0 (0.0\%) \\
6+ & 353 (2.2\%) & 280 (1.7\%) & 216 (1.3\%) & 189 (1.1\%) & 0 (0.0\%) \\
\multicolumn{6}{@{}l}{\textbf{Household Wealth Index}} \\
Mean (sd) & 2.5 (0.46) & 2.5 (0.46) & 2.6 (0.47) & 2.6 (0.47) & 2.6 (0.44) \\
Median {[}Min - Max{]} & 2.6 {[}0.90 - 4.0{]} & 2.6 {[}0.90 - 4.0{]} & 2.7 {[}0.82 - 4.1{]} & 2.7 {[}0.90 - 4.1{]} & 2.7 {[}0.94 - 4.2{]} \\
Missing (\%) & 1498 (9.2\%) & 1517 (9.1\%) & 3557 (20.9\%) & 3539 (20.2\%) & 4631 (26.0\%) \\
\multicolumn{6}{@{}l}{\textbf{Network Wealth Index}} \\
Mean (sd) & 5.3 (5.7) & 5.4 (5.7) & 5.7 (6.0) & 5.7 (6.0) & 5.8 (6.2) \\
Median {[}Min - Max{]} & 3.8 {[}0.00 - 46{]} & 4.1 {[}0.00 - 47{]} & 4.5 {[}0.00 - 45{]} & 4.4 {[}0.00 - 56{]} & 4.7 {[}0.00 - 55{]} \\
Missing (\%) & 262 (1.6\%) & 238 (1.4\%) & 592 (3.5\%) & 532 (3.0\%) & 863 (4.8\%) \\
\end{longtable}
}

\textbf{\hfill\break
}

{\footnotesize\setlength{\tabcolsep}{2.5pt}
\begin{longtable}[]{@{}lrrrrr@{}}
\toprule\noalign{}
 & 2014 & 2015 & 2016 & 2017 & 2018 \\
\midrule\noalign{}
\endhead
\bottomrule\noalign{}
\endlastfoot
\multicolumn{6}{@{}l}{\textbf{Number of Households}} \\
N & 18144 (100.0\%) & 18332 (100.0\%) & 18461 (100.0\%) & 18337 (100.0\%) & 17924 (100.0\%) \\
\multicolumn{6}{@{}l}{\textbf{Village Type}} \\
Original Villages & 18144 (100.0\%) & 18332 (100.0\%) & 18461 (100.0\%) & 18337 (100.0\%) & 17924 (100.0\%) \\
\multicolumn{6}{@{}l}{\textbf{\# Not in Employment Market}} \\
0 & 18137 (100.0\%) & 18322 (99.9\%) & 4267 (23.1\%) & 4302 (23.5\%) & 4222 (23.6\%) \\
1 & 7 (0.0\%) & 10 (0.1\%) & 4822 (26.1\%) & 4838 (26.4\%) & 4792 (26.7\%) \\
2 & 0 (0.0\%) & 0 (0.0\%) & 3961 (21.5\%) & 3955 (21.6\%) & 3897 (21.7\%) \\
3 & 0 (0.0\%) & 0 (0.0\%) & 2575 (13.9\%) & 2545 (13.9\%) & 2498 (13.9\%) \\
4 & 0 (0.0\%) & 0 (0.0\%) & 1511 (8.2\%) & 1444 (7.9\%) & 1373 (7.7\%) \\
5 & 0 (0.0\%) & 0 (0.0\%) & 752 (4.1\%) & 725 (4.0\%) & 657 (3.7\%) \\
6+ & 0 (0.0\%) & 0 (0.0\%) & 573 (3.1\%) & 528 (2.9\%) & 485 (2.7\%) \\
\multicolumn{6}{@{}l}{\textbf{\# Not Working}} \\
0 & 18124 (99.9\%) & 18304 (99.8\%) & 2356 (12.8\%) & 1989 (10.8\%) & 1782 (9.9\%) \\
1 & 19 (0.1\%) & 27 (0.1\%) & 3928 (21.3\%) & 3740 (20.4\%) & 3060 (17.1\%) \\
2 & 1 (0.0\%) & 1 (0.0\%) & 3754 (20.3\%) & 3714 (20.3\%) & 3334 (18.6\%) \\
3 & 0 (0.0\%) & 0 (0.0\%) & 3051 (16.5\%) & 3076 (16.8\%) & 3131 (17.5\%) \\
4 & 0 (0.0\%) & 0 (0.0\%) & 2110 (11.4\%) & 2278 (12.4\%) & 2453 (13.7\%) \\
5 & 0 (0.0\%) & 0 (0.0\%) & 1472 (8.0\%) & 1523 (8.3\%) & 1617 (9.0\%) \\
6+ & 0 (0.0\%) & 0 (0.0\%) & 1790 (9.7\%) & 2017 (11.0\%) & 2547 (14.2\%) \\
\multicolumn{6}{@{}l}{\textbf{\# Currently Working}} \\
0 & 18135 (100.0\%) & 18318 (99.9\%) & 4904 (26.6\%) & 5275 (28.8\%) & 4324 (24.1\%) \\
1 & 9 (0.0\%) & 14 (0.1\%) & 7550 (40.9\%) & 7487 (40.8\%) & 7238 (40.4\%) \\
2 & 0 (0.0\%) & 0 (0.0\%) & 3780 (20.5\%) & 3554 (19.4\%) & 3834 (21.4\%) \\
3 & 0 (0.0\%) & 0 (0.0\%) & 1405 (7.6\%) & 1275 (7.0\%) & 1565 (8.7\%) \\
4 & 0 (0.0\%) & 0 (0.0\%) & 540 (2.9\%) & 507 (2.8\%) & 621 (3.5\%) \\
5 & 0 (0.0\%) & 0 (0.0\%) & 190 (1.0\%) & 159 (0.9\%) & 214 (1.2\%) \\
6+ & 0 (0.0\%) & 0 (0.0\%) & 92 (0.5\%) & 80 (0.4\%) & 128 (0.7\%) \\
\multicolumn{6}{@{}l}{\textbf{\# Unemployed}} \\
0 & 18121 (99.9\%) & 18299 (99.8\%) & 2024 (11.0\%) & 2023 (11.0\%) & 1995 (11.1\%) \\
1 & 23 (0.1\%) & 33 (0.2\%) & 5681 (30.8\%) & 5848 (31.9\%) & 5871 (32.8\%) \\
2 & 0 (0.0\%) & 0 (0.0\%) & 5177 (28.0\%) & 5138 (28.0\%) & 5031 (28.1\%) \\
3 & 0 (0.0\%) & 0 (0.0\%) & 2681 (14.5\%) & 2606 (14.2\%) & 2535 (14.1\%) \\
4 & 0 (0.0\%) & 0 (0.0\%) & 1529 (8.3\%) & 1453 (7.9\%) & 1368 (7.6\%) \\
5 & 0 (0.0\%) & 0 (0.0\%) & 753 (4.1\%) & 729 (4.0\%) & 655 (3.7\%) \\
6+ & 0 (0.0\%) & 0 (0.0\%) & 616 (3.3\%) & 540 (2.9\%) & 469 (2.6\%) \\
\multicolumn{6}{@{}l}{\textbf{Household Wealth Index}} \\
Mean (sd) & 2.6 (0.41) & 2.6 (0.40) & 2.7 (0.38) & 2.6 (0.39) & 2.7 (0.39) \\
Median {[}Min - Max{]} & 2.7 {[}0.90 - 4.0{]} & 2.7 {[}0.92 - 3.9{]} & 2.7 {[}0.90 - 4.0{]} & 2.7 {[}0.86 - 4.0{]} & 2.7 {[}0.92 - 4.2{]} \\
Missing (\%) & 3672 (20.2\%) & 4198 (22.9\%) & 4229 (22.9\%) & 6443 (35.1\%) & 4033 (22.5\%) \\
\multicolumn{6}{@{}l}{\textbf{Network Wealth Index}} \\
Mean (sd) & 5.9 (6.2) & 6.0 (6.3) & 6.0 (6.3) & 6.0 (6.3) & 5.7 (6.0) \\
Median {[}Min - Max{]} & 4.8 {[}0.00 - 52{]} & 4.9 {[}0.00 - 55{]} & 5.0 {[}0.00 - 52{]} & 4.9 {[}0.00 - 54{]} & 4.7 {[}0.00 - 57{]} \\
Missing (\%) & 634 (3.5\%) & 769 (4.2\%) & 800 (4.3\%) & 1631 (8.9\%) & 894 (5.0\%) \\
\end{longtable}
}

\hypertarget{supp-section-4}{%
\subsubsection{\texorpdfstring{\hfill\break
}{ }}\label{supp-section-4}}

\hypertarget{supp-additional-villages-demographic-variables}{%
\subsection{Additional Villages Demographic Variables}\label{supp-additional-villages-demographic-variables}}

{\footnotesize\setlength{\tabcolsep}{2.5pt}
\begin{longtable}[]{@{}lrrrrrr@{}}
\toprule\noalign{}
 & 2013 & 2014 & 2015 & 2016 & 2017 & 2018 \\
\midrule\noalign{}
\endhead
\bottomrule\noalign{}
\endlastfoot
\multicolumn{7}{@{}l}{\textbf{Number of Households}} \\
N & 3294 (100.0\%) & 3383 (100.0\%) & 3476 (100.0\%) & 3487 (100.0\%) & 3451 (100.0\%) & 3367 (100.0\%) \\
\multicolumn{7}{@{}l}{\textbf{Village Type}} \\
Additional Villages & 3294 (100.0\%) & 3383 (100.0\%) & 3476 (100.0\%) & 3487 (100.0\%) & 3451 (100.0\%) & 3367 (100.0\%) \\
\multicolumn{7}{@{}l}{\textbf{\# Connected Households}} \\
0 & 2743 (83.3\%) & 2675 (79.1\%) & 2635 (75.8\%) & 2587 (74.2\%) & 2565 (74.3\%) & 2538 (75.4\%) \\
1 & 428 (13.0\%) & 530 (15.7\%) & 597 (17.2\%) & 625 (17.9\%) & 621 (18.0\%) & 588 (17.5\%) \\
2 & 91 (2.8\%) & 126 (3.7\%) & 167 (4.8\%) & 184 (5.3\%) & 181 (5.2\%) & 168 (5.0\%) \\
3 & 16 (0.5\%) & 30 (0.9\%) & 50 (1.4\%) & 53 (1.5\%) & 46 (1.3\%) & 42 (1.2\%) \\
4 & 5 (0.2\%) & 6 (0.2\%) & 7 (0.2\%) & 18 (0.5\%) & 18 (0.5\%) & 17 (0.5\%) \\
5 & 9 (0.3\%) & 9 (0.3\%) & 13 (0.4\%) & 8 (0.2\%) & 7 (0.2\%) & 5 (0.1\%) \\
6+ & 2 (0.1\%) & 7 (0.2\%) & 7 (0.2\%) & 12 (0.3\%) & 13 (0.4\%) & 9 (0.3\%) \\
\multicolumn{7}{@{}l}{\textbf{\# Household Members}} \\
1 & 349 (10.6\%) & 362 (10.7\%) & 384 (11.0\%) & 370 (10.6\%) & 366 (10.6\%) & 365 (10.8\%) \\
2-5 & 1472 (44.7\%) & 1520 (44.9\%) & 1577 (45.4\%) & 1597 (45.8\%) & 1584 (45.9\%) & 1553 (46.1\%) \\
6-10 & 1132 (34.4\%) & 1158 (34.2\%) & 1161 (33.4\%) & 1166 (33.4\%) & 1155 (33.5\%) & 1129 (33.5\%) \\
11-20 & 331 (10.0\%) & 331 (9.8\%) & 343 (9.9\%) & 342 (9.8\%) & 333 (9.6\%) & 312 (9.3\%) \\
21+ & 10 (0.3\%) & 12 (0.4\%) & 11 (0.3\%) & 12 (0.3\%) & 13 (0.4\%) & 8 (0.2\%) \\
\multicolumn{7}{@{}l}{\textbf{\# Children (\textless5)}} \\
0 & 1787 (54.3\%) & 1826 (54.0\%) & 1935 (55.7\%) & 2008 (57.6\%) & 2061 (59.7\%) & 2117 (62.9\%) \\
1 & 995 (30.2\%) & 1054 (31.2\%) & 1041 (29.9\%) & 1015 (29.1\%) & 986 (28.6\%) & 920 (27.3\%) \\
2-5 & 509 (15.5\%) & 500 (14.8\%) & 497 (14.3\%) & 462 (13.2\%) & 401 (11.6\%) & 330 (9.8\%) \\
6-10 & 3 (0.1\%) & 3 (0.1\%) & 3 (0.1\%) & 2 (0.1\%) & 3 (0.1\%) & 0 (0.0\%) \\
\multicolumn{7}{@{}l}{\textbf{\# Adults (15-59)}} \\
0 & 126 (3.8\%) & 121 (3.6\%) & 130 (3.7\%) & 127 (3.6\%) & 120 (3.5\%) & 125 (3.7\%) \\
1 & 474 (14.4\%) & 498 (14.7\%) & 491 (14.1\%) & 505 (14.5\%) & 510 (14.8\%) & 488 (14.5\%) \\
2-5 & 2099 (63.7\%) & 2168 (64.1\%) & 2247 (64.6\%) & 2254 (64.6\%) & 2233 (64.7\%) & 2192 (65.1\%) \\
6-10 & 554 (16.8\%) & 559 (16.5\%) & 581 (16.7\%) & 571 (16.4\%) & 557 (16.1\%) & 538 (16.0\%) \\
11-20 & 41 (1.2\%) & 37 (1.1\%) & 27 (0.8\%) & 30 (0.9\%) & 31 (0.9\%) & 24 (0.7\%) \\
\multicolumn{7}{@{}l}{\textbf{\# Elders (60+)}} \\
0 & 2286 (69.4\%) & 2345 (69.3\%) & 2408 (69.3\%) & 2391 (68.6\%) & 2336 (67.7\%) & 2196 (65.2\%) \\
1 & 800 (24.3\%) & 824 (24.4\%) & 843 (24.3\%) & 873 (25.0\%) & 887 (25.7\%) & 922 (27.4\%) \\
2 & 203 (6.2\%) & 208 (6.1\%) & 216 (6.2\%) & 214 (6.1\%) & 221 (6.4\%) & 244 (7.2\%) \\
3+ & 5 (0.2\%) & 6 (0.2\%) & 9 (0.3\%) & 9 (0.3\%) & 7 (0.2\%) & 5 (0.1\%) \\
\multicolumn{7}{@{}l}{\textbf{\# Adult Men}} \\
0 & 541 (16.4\%) & 534 (15.8\%) & 539 (15.5\%) & 549 (15.7\%) & 538 (15.6\%) & 514 (15.3\%) \\
1 & 1290 (39.2\%) & 1331 (39.3\%) & 1386 (39.9\%) & 1373 (39.4\%) & 1336 (38.7\%) & 1291 (38.3\%) \\
2 & 723 (21.9\%) & 755 (22.3\%) & 784 (22.6\%) & 786 (22.5\%) & 790 (22.9\%) & 793 (23.6\%) \\
3 & 438 (13.3\%) & 432 (12.8\%) & 421 (12.1\%) & 437 (12.5\%) & 445 (12.9\%) & 434 (12.9\%) \\
4 & 174 (5.3\%) & 196 (5.8\%) & 219 (6.3\%) & 211 (6.1\%) & 217 (6.3\%) & 212 (6.3\%) \\
5 & 80 (2.4\%) & 87 (2.6\%) & 82 (2.4\%) & 88 (2.5\%) & 81 (2.3\%) & 87 (2.6\%) \\
6+ & 48 (1.5\%) & 48 (1.4\%) & 45 (1.3\%) & 43 (1.2\%) & 44 (1.3\%) & 36 (1.1\%) \\
\multicolumn{7}{@{}l}{\textbf{\# Adult Women}} \\
0 & 519 (15.8\%) & 539 (15.9\%) & 550 (15.8\%) & 555 (15.9\%) & 551 (16.0\%) & 573 (17.0\%) \\
1 & 1151 (34.9\%) & 1186 (35.1\%) & 1238 (35.6\%) & 1233 (35.4\%) & 1208 (35.0\%) & 1142 (33.9\%) \\
2 & 729 (22.1\%) & 748 (22.1\%) & 772 (22.2\%) & 782 (22.4\%) & 809 (23.4\%) & 803 (23.8\%) \\
3 & 458 (13.9\%) & 492 (14.5\%) & 476 (13.7\%) & 478 (13.7\%) & 456 (13.2\%) & 446 (13.2\%) \\
4 & 246 (7.5\%) & 232 (6.9\%) & 254 (7.3\%) & 255 (7.3\%) & 246 (7.1\%) & 241 (7.2\%) \\
5 & 119 (3.6\%) & 114 (3.4\%) & 119 (3.4\%) & 118 (3.4\%) & 114 (3.3\%) & 97 (2.9\%) \\
6+ & 72 (2.2\%) & 72 (2.1\%) & 67 (1.9\%) & 66 (1.9\%) & 67 (1.9\%) & 65 (1.9\%) \\
\multicolumn{7}{@{}l}{\textbf{\# Elder Men}} \\
0 & 2861 (86.9\%) & 2945 (87.1\%) & 3016 (86.8\%) & 3008 (86.3\%) & 2965 (85.9\%) & 2853 (84.7\%) \\
1 & 432 (13.1\%) & 436 (12.9\%) & 458 (13.2\%) & 475 (13.6\%) & 484 (14.0\%) & 510 (15.1\%) \\
2 & 1 (0.0\%) & 2 (0.1\%) & 2 (0.1\%) & 4 (0.1\%) & 2 (0.1\%) & 4 (0.1\%) \\
\multicolumn{7}{@{}l}{\textbf{\# Elder Women}} \\
0 & 2535 (77.0\%) & 2590 (76.6\%) & 2662 (76.6\%) & 2664 (76.4\%) & 2610 (75.6\%) & 2484 (73.8\%) \\
1 & 732 (22.2\%) & 769 (22.7\%) & 789 (22.7\%) & 802 (23.0\%) & 821 (23.8\%) & 860 (25.5\%) \\
2 & 26 (0.8\%) & 23 (0.7\%) & 23 (0.7\%) & 19 (0.5\%) & 18 (0.5\%) & 21 (0.6\%) \\
3 & 1 (0.0\%) & 1 (0.0\%) & 2 (0.1\%) & 2 (0.1\%) & 2 (0.1\%) & 2 (0.1\%) \\
\multicolumn{7}{@{}l}{\textbf{\# Deaths in Household}} \\
0 & 3187 (96.8\%) & 3260 (96.4\%) & 3328 (95.7\%) & 3352 (96.1\%) & 3338 (96.7\%) & 3284 (97.5\%) \\
1 & 103 (3.1\%) & 118 (3.5\%) & 145 (4.2\%) & 131 (3.8\%) & 109 (3.2\%) & 81 (2.4\%) \\
2 & 4 (0.1\%) & 5 (0.1\%) & 3 (0.1\%) & 4 (0.1\%) & 3 (0.1\%) & 2 (0.1\%) \\
3+ & 0 (0.0\%) & 0 (0.0\%) & 0 (0.0\%) & 0 (0.0\%) & 1 (0.0\%) & 0 (0.0\%) \\
\end{longtable}
}

\hypertarget{supp-additional-villages-economic-variables}{%
\subsection{Additional Villages Economic Variables}\label{supp-additional-villages-economic-variables}}

{\footnotesize\setlength{\tabcolsep}{2.5pt}
\begin{longtable}[]{@{}lrrrrrr@{}}
\toprule\noalign{}
 & 2013 & 2014 & 2015 & 2016 & 2017 & 2018 \\
\midrule\noalign{}
\endhead
\bottomrule\noalign{}
\endlastfoot
\multicolumn{7}{@{}l}{\textbf{Number of Households}} \\
N & 3294 (100.0\%) & 3383 (100.0\%) & 3476 (100.0\%) & 3487 (100.0\%) & 3451 (100.0\%) & 3367 (100.0\%) \\
\multicolumn{7}{@{}l}{\textbf{Village Type}} \\
Additional Villages & 3294 (100.0\%) & 3383 (100.0\%) & 3476 (100.0\%) & 3487 (100.0\%) & 3451 (100.0\%) & 3367 (100.0\%) \\
\multicolumn{7}{@{}l}{\textbf{\# Not in Employment Market}} \\
0 & 1226 (37.2\%) & 1338 (39.6\%) & 1462 (42.1\%) & 741 (21.3\%) & 749 (21.7\%) & 745 (22.1\%) \\
1 & 1020 (31.0\%) & 1046 (30.9\%) & 1066 (30.7\%) & 894 (25.6\%) & 883 (25.6\%) & 860 (25.5\%) \\
2 & 617 (18.7\%) & 607 (17.9\%) & 583 (16.8\%) & 740 (21.2\%) & 730 (21.2\%) & 725 (21.5\%) \\
3 & 277 (8.4\%) & 258 (7.6\%) & 246 (7.1\%) & 521 (14.9\%) & 523 (15.2\%) & 512 (15.2\%) \\
4 & 98 (3.0\%) & 83 (2.5\%) & 79 (2.3\%) & 298 (8.5\%) & 291 (8.4\%) & 273 (8.1\%) \\
5 & 39 (1.2\%) & 36 (1.1\%) & 28 (0.8\%) & 165 (4.7\%) & 155 (4.5\%) & 145 (4.3\%) \\
6+ & 17 (0.5\%) & 15 (0.4\%) & 12 (0.3\%) & 128 (3.7\%) & 120 (3.5\%) & 107 (3.2\%) \\
\multicolumn{7}{@{}l}{\textbf{\# Not Working}} \\
0 & 584 (17.7\%) & 665 (19.7\%) & 751 (21.6\%) & 434 (12.4\%) & 854 (24.7\%) & 330 (9.8\%) \\
1 & 819 (24.9\%) & 878 (26.0\%) & 944 (27.2\%) & 681 (19.5\%) & 608 (17.6\%) & 572 (17.0\%) \\
2 & 683 (20.7\%) & 687 (20.3\%) & 697 (20.1\%) & 643 (18.4\%) & 531 (15.4\%) & 626 (18.6\%) \\
3 & 487 (14.8\%) & 504 (14.9\%) & 481 (13.8\%) & 568 (16.3\%) & 482 (14.0\%) & 545 (16.2\%) \\
4 & 331 (10.0\%) & 310 (9.2\%) & 296 (8.5\%) & 434 (12.4\%) & 358 (10.4\%) & 428 (12.7\%) \\
5 & 203 (6.2\%) & 193 (5.7\%) & 184 (5.3\%) & 318 (9.1\%) & 244 (7.1\%) & 315 (9.4\%) \\
6+ & 187 (5.7\%) & 146 (4.3\%) & 123 (3.5\%) & 409 (11.7\%) & 374 (10.8\%) & 551 (16.4\%) \\
\multicolumn{7}{@{}l}{\textbf{\# Currently Working}} \\
0 & 1078 (32.7\%) & 1169 (34.6\%) & 1270 (36.5\%) & 1020 (29.3\%) & 1338 (38.8\%) & 725 (21.5\%) \\
1 & 1226 (37.2\%) & 1263 (37.3\%) & 1298 (37.3\%) & 1264 (36.2\%) & 1101 (31.9\%) & 1302 (38.7\%) \\
2 & 588 (17.9\%) & 580 (17.1\%) & 554 (15.9\%) & 724 (20.8\%) & 602 (17.4\%) & 742 (22.0\%) \\
3 & 226 (6.9\%) & 216 (6.4\%) & 219 (6.3\%) & 293 (8.4\%) & 254 (7.4\%) & 336 (10.0\%) \\
4 & 116 (3.5\%) & 105 (3.1\%) & 90 (2.6\%) & 105 (3.0\%) & 97 (2.8\%) & 162 (4.8\%) \\
5 & 41 (1.2\%) & 33 (1.0\%) & 31 (0.9\%) & 54 (1.5\%) & 40 (1.2\%) & 64 (1.9\%) \\
6+ & 19 (0.6\%) & 17 (0.5\%) & 14 (0.4\%) & 27 (0.8\%) & 19 (0.6\%) & 36 (1.1\%) \\
\multicolumn{7}{@{}l}{\textbf{\# Unemployed}} \\
0 & 447 (13.6\%) & 514 (15.2\%) & 584 (16.8\%) & 433 (12.4\%) & 430 (12.5\%) & 434 (12.9\%) \\
1 & 754 (22.9\%) & 816 (24.1\%) & 889 (25.6\%) & 984 (28.2\%) & 1009 (29.2\%) & 996 (29.6\%) \\
2 & 944 (28.7\%) & 964 (28.5\%) & 967 (27.8\%) & 889 (25.5\%) & 886 (25.7\%) & 870 (25.8\%) \\
3 & 476 (14.5\%) & 468 (13.8\%) & 465 (13.4\%) & 521 (14.9\%) & 508 (14.7\%) & 491 (14.6\%) \\
4 & 283 (8.6\%) & 287 (8.5\%) & 269 (7.7\%) & 300 (8.6\%) & 288 (8.3\%) & 290 (8.6\%) \\
5 & 196 (6.0\%) & 180 (5.3\%) & 177 (5.1\%) & 178 (5.1\%) & 161 (4.7\%) & 142 (4.2\%) \\
6+ & 194 (5.9\%) & 154 (4.6\%) & 125 (3.6\%) & 182 (5.2\%) & 169 (4.9\%) & 144 (4.3\%) \\
\multicolumn{7}{@{}l}{\textbf{Household Wealth Index}} \\
Mean (sd) & 2.7 (0.47) & 2.7 (0.42) & 2.8 (0.38) & 2.8 (0.36) & 2.8 (0.38) & 2.8 (0.38) \\
Median {[}Min - Max{]} & 2.7 {[}0.92 - 4.2{]} & 2.8 {[}1.2 - 3.9{]} & 2.8 {[}1.2 - 4.0{]} & 2.8 {[}1.2 - 4.3{]} & 2.8 {[}1.2 - 3.9{]} & 2.8 {[}0.95 - 4.0{]} \\
Missing (\%) & 866 (26.3\%) & 668 (19.7\%) & 641 (18.4\%) & 860 (24.7\%) & 892 (25.8\%) & 728 (21.6\%) \\
\multicolumn{7}{@{}l}{\textbf{Network Wealth Index}} \\
Mean (sd) & 0.51 (1.5) & 0.72 (1.9) & 0.88 (2.1) & 0.98 (2.3) & 0.94 (2.2) & 0.89 (2.1) \\
Median {[}Min - Max{]} & 0.00 {[}0.00 - 15{]} & 0.00 {[}0.00 - 33{]} & 0.00 {[}0.00 - 29{]} & 0.00 {[}0.00 - 26{]} & 0.00 {[}0.00 - 27{]} & 0.00 {[}0.00 - 26{]} \\
Missing (\%) & 111 (3.4\%) & 110 (3.3\%) & 132 (3.8\%) & 153 (4.4\%) & 171 (5.0\%) & 131 (3.9\%) \\
\end{longtable}
}

\hypertarget{supp-appendix-s4-causal-estimands}{%
\section{Appendix S4: Causal Estimands}\label{supp-appendix-s4-causal-estimands}}

Direct and Spillover Effect of Family Network Wealth on Household Death. Agincourt Health and Demographic Surveillance System: 2009 to 2018.

{\footnotesize\setlength{\tabcolsep}{2.5pt}
\begin{longtable}[]{@{}
  >{\raggedright\arraybackslash}p{(\columnwidth - 6\tabcolsep) * \real{0.1815}}
  >{\raggedright\arraybackslash}p{(\columnwidth - 6\tabcolsep) * \real{0.2838}}
  >{\raggedright\arraybackslash}p{(\columnwidth - 6\tabcolsep) * \real{0.2613}}
  >{\raggedright\arraybackslash}p{(\columnwidth - 6\tabcolsep) * \real{0.2735}}@{}}
\caption*{Table S2: Causal effects of household wealth on household death -- multiple imputation analysis}\tabularnewline
\toprule\noalign{}
 & \textbf{ADE}

Est. (95\%CI) & \textbf{ASE}

Est. (95\%CI) & \textbf{ATE}

Est. (95\%CI) \\
\midrule\noalign{}
\endfirsthead
\toprule\noalign{}
 & \textbf{ADE}

Est. (95\%CI) & \textbf{ASE}

Est. (95\%CI) & \textbf{ATE}

Est. (95\%CI) \\
\midrule\noalign{}
\endhead
\bottomrule\noalign{}
\endlastfoot
\textbf{Children} & -0.15\%

(-0.19 to -0.11)

P\textless0.001 & -0.02\%

(-0.05 to 0.02)

P=0.328 & -0.17\%

(-0.22 to -0.12)

P\textless0.001 \\
\textbf{Adults} & -0.29\%

(-0.37 to -0.21)

P\textless0.001 & -0.08\%

(-0.15 to 0.00)

P=0.038 & -0.37\%

(-0.47 to -0.26)

P\textless0.001 \\
\textbf{Elders} & -0.09\%

(-0.14 to -0.03)

P=0.002 & 0.01\%

(-0.04 to 0.06)

P=0.758 & -0.08\%

(-0.15 to -0.01)

P=0.029 \\
\textbf{Overall} & -0.53\%

(-0.63 to -0.42)

P\textless0.001 & -0.08\%

(-0.18 to 0.01)

P=0.075 & -0.61\%

(-0.74 to -0.48)

P\textless0.001 \\
\end{longtable}
}

\end{document}